\documentclass[twocolumn,aps,floatfix,amsmath,amssymb,superscriptaddress,nofootinbib,longbibliography]{revtex4-2}

\usepackage{graphicx}
\usepackage{color}
\usepackage{soul}
\usepackage{bm}
\usepackage{amsmath}
\usepackage{amssymb}
\usepackage{amsthm}
\usepackage{amsfonts}
\usepackage{verbatim}
\usepackage{float}
\usepackage[caption=false]{subfig}
\usepackage[dvipsnames]{xcolor}

\usepackage{hyperref}

\begin{document}

\title{Experimental observation of exceptional bound states in a classical circuit network}

\author{Deyuan Zou}
\thanks{These authors contributed equally to this work.}
\affiliation{Key Laboratory of Advanced Optoelectronic Quantum Architecture and Measurements of Ministry of Education, Beijing Key Laboratory of Nanophotonics and Ultrafine Optoelectronic Systems, School of Physics, Beijing Institute of Technology, Beijing 100081, China.}

\author{Tian Chen}
\thanks{These authors contributed equally to this work.}
\affiliation{Key Laboratory of Advanced Optoelectronic Quantum Architecture and Measurements of Ministry of Education, Beijing Key Laboratory of Nanophotonics and Ultrafine Optoelectronic Systems, School of Physics, Beijing Institute of Technology, Beijing 100081, China.}

\author{Haiyu Meng}
\email{phymhy@xtu.edu.cn}
\affiliation{School of Physics and Optoelectronics, Xiangtan University, Xiangtan 411105, China}
\affiliation{Department of Physics, National University of Singapore, Singapore 117542}

\author{Yee Sin Ang}
\email{yeesin$_$ang@sutd.edu.sg}
\affiliation{Science, Mathematics and Technology, Singapore University of Technology and Design, Singapore 487372}

\author{Xiangdong Zhang}
\email{zhangxd@bit.edu.cn}
\affiliation{Key Laboratory of Advanced Optoelectronic Quantum Architecture and Measurements of Ministry of Education, Beijing Key Laboratory of Nanophotonics and Ultrafine Optoelectronic Systems, School of Physics, Beijing Institute of Technology, Beijing 100081, China.}

\author{Ching Hua Lee}
\email{phylch@nus.edu.sg}
\affiliation{Department of Physics, National University of Singapore, Singapore 117542}
\affiliation{Joint School of National University of Singapore and Tianjin University, International Campus of Tianjin University, Binhai New City, Fuzhou 350207, China}


\maketitle
\date{\today}


\textbf{
Exceptional bound (EB) states represent an unique new class of robust bound states protected by the defectiveness of non-Hermitian exceptional points. Conceptually distinct from the more well-known topological states and non-Hermitian skin states, they were recently discovered~\cite{lee2022exceptional} as a novel source of negative entanglement entropy in the quantum entanglement context. Yet, EB states have been physically elusive, being originally interpreted as negative probability eigenstates of the propagator of non-Hermitian Fermi gases. In this work, we show that EB states are in fact far more ubiquitous, also arising robustly in broad classes of systems whether classical or quantum. 
This hinges crucially on a newly-discovered spectral flow that rigorously justifies the EB nature of small candidate lattice systems. As a highlight, we present their first experimental realization through an electrical circuit, where they manifest as prominent stable resonant voltage profiles. Our work brings a hitherto elusive but fundamentally distinctive quantum phenomenon into the realm of classical metamaterials, and provides a novel pathway for the engineering of robust modes in otherwise sensitive systems.
}

In 2020, it was first realized that negative entanglement entropy can mysteriously arise in a non-Hermitian free fermion lattice~\cite{lee2022exceptional}. First attributed to non-unitary bc-ghost conformal field theory (CFT) by Ryu et al~\cite{chang2020entanglement}, it was subsequently realized~\cite{lee2022exceptional} that it generically occurs in the presence of exceptional points, which are branch points in the complex bandstructure~\cite{dembowski2004encircling,rotter2009non,jin2009solutions,longhi2010pt,heiss2001chirality,heiss2012physics,hassan2017dynamically,xu2016topological,lin2017line,hu2017exceptional,shen2018topological,wang2019arbitrary,miri2019exceptional,zhang2020non,jin2020hybrid,kawabata2019classification}. While exceptional points have been been extensively theoretically investigated and experimentally observed in a variety of classical metamaterials~\cite{feng2013experimental, liu2018unidirectional}, what was not previously known is that its defectiveness i.e. incompleteness of eigenbasis also has profound impact on its quantum information aspects. Specifically, the defectiveness allows the free fermion propagator to become divergently asymmetric, leading to a non-cancellation of particle hoppings across an entanglement cut, reminiscent of Hawking radiation near the surface of a black hole~\cite{almheiri2020entropy,chen2020quantum}. This non-conservation of probability leads to negative entanglement.

A very pertinent question is: can such esoteric negative entanglement EB state behavior manifest in realistic experiments? In this work, we theoretically and experimentally provide an affirmative ``yes''. As a special type of mathematical eigenstate, EB states are fundamentally not restricted to any particular physical setup. In particular, even though EB states are originally defined as negative probability eigenstates of a fermion propagator~\cite{lee2022exceptional}, they can equivalently exist as physical eigenstates of network Laplacians with an identical mathematical form.

Key to the viability of EB states in classical networks is their remarkable robustness. It is commonly believed that highly non-local operators, such as the parent propagators of EB states, cannot be physically realized as a lattice model due to the prohibitively large number of effective hoppings. 
In this work, we discover a new form of adiabatic spectral continuity of negative probability eigenvalues, reminiscent of the (completely physically different) spectral flow due to topological pumping~\cite{qi2011generic,alexandradinata2011trace,yu2011equivalent,lee2015free}, that ensures that EB states continue to survive almost undisturbed even when the parent propagator matrix is heavily truncated. Indeed, EB states remain prominently well-gapped in very small lattice networks down to the order of $\mathcal{O}(10^1)$ sites, 
greatly facilitating their physical realization in relatively simple classical networks such as electrical circuits without the need for fine-tuning.

As the highlight, we experimentally implemented a defective free-fermion propagator matrix in an electrical circuit network, and observed resonant voltage profiles that exhibits excellent agreement with those of theoretically predicted EB states. While the success of this experimental demonstration is aided by the versatility of circuit connections, it more crucially hinges on the abovementioned newly-discovered spectral flow that enables the rigorous preservation of EB states in sufficiently small lattices with manageable numbers of non-local couplings.

\section*{Results}
\label{sec:results}

\subsection{Exceptional bound (EB) states as a novel type of robustness}
\label{sec:EBintro}

Robust bound states are highly touted for their applications in topological waveguides, non-Hermitian sensing or BIC systems~\cite{kim2021quantum, weimann2017topologically, budich2020non, hsu2016bound, zhen2014topological,morvan2022formation}. Recent literature features two conventional routes for generating robust bound states - (i) topology and (ii) the non-Hermitian skin effect (NHSE). In the former, robust boundary-localized states are protected by bulk topological invariants such as the Chern or $\mathbb{Z}_2$ winding numbers~\cite{haldane1988model,kane2005z,schnyder2009classification,chang2023colloquium}. 
In the latter, ``skin'' states accumulate robustly at system boundaries due to the directional amplification from asymmetric hoppings~\cite{yao2018edge,xiong2018does,lee2019anatomy,yokomizo2019non,kunst2018biorthogonal,lee2020unraveling}.

\begin{figure}[!t]
    \centering
    \includegraphics[width =\linewidth]{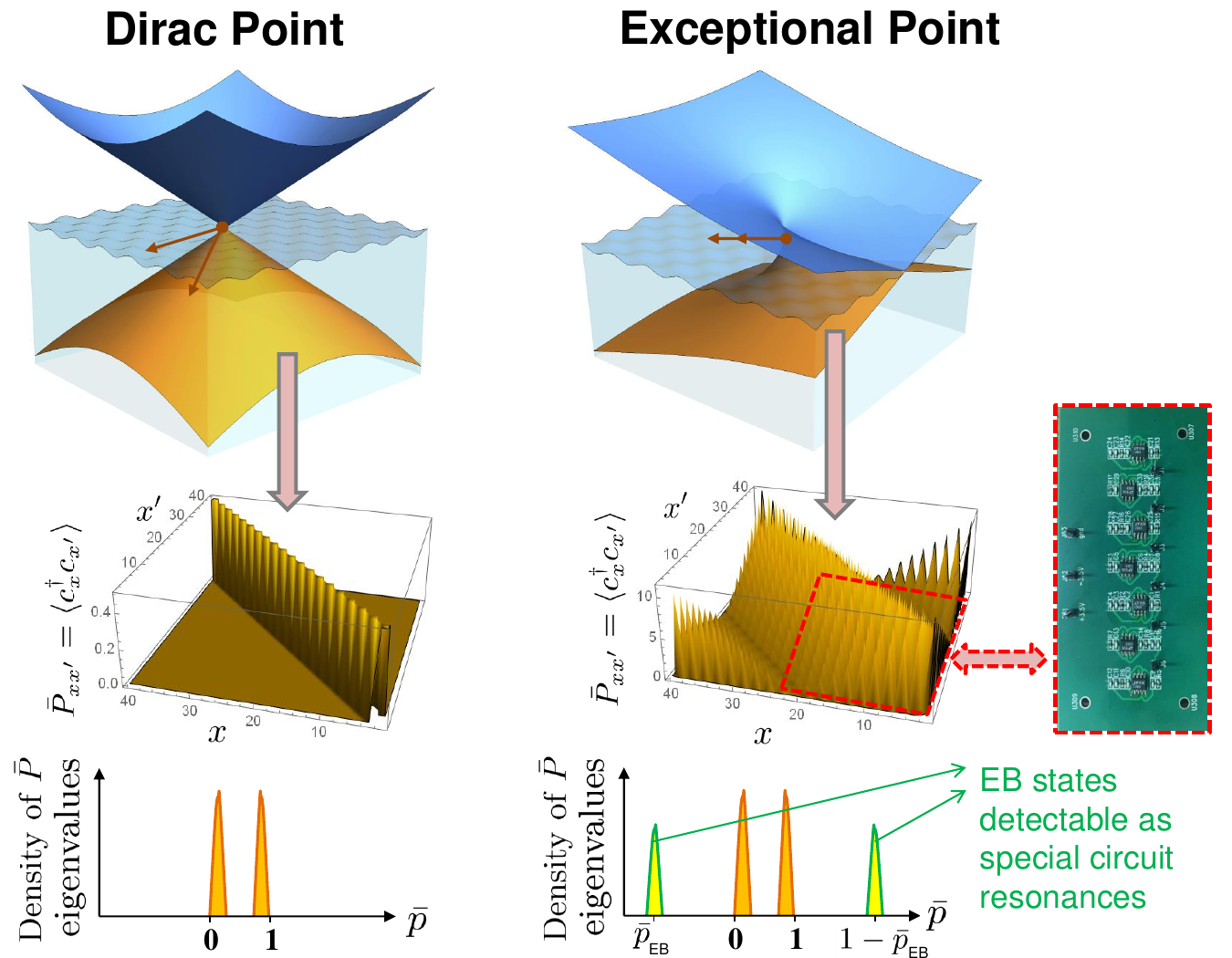}
    \caption{\textbf{Origin of EB states.} In the traditional quantum entanglement context, gapless points (Top) are characterized by their bounded occupied Fermi sea projector $\bar P$ (Middle) whose eigenvalues $\bar p$ indicates the occupancy of the lower band (orange). While usual gapless points that are not geometrically defective i.e. Dirac points (Left column) possess only $\bar P$ eigenvalues within $[0,1]$ (Bottom Left), defective exceptional points (Right column) also exhibit special isolated EB eigenvalues $\bar p_\text{EB}$, $1-\bar p_\text{EB}$ far outside of $[0,1]$ (Bottom Right). The enigmatic negative occupation $\bar p_\text{EB}<0 $ stems from the non-locality of the 2-point functions that form the matrix elements of $\bar P$ (Middle Right), and can be detected as pronounced impedance resonances if $\bar P$, which is highly non-local, were to be implemented as the Laplacian of an electric circuit.
}
    \label{fig:1}
\end{figure}

By contrast, in this work, we theoretically and experimentally characterize a new type of robustness from so-called exceptional bound (EB) states, whose robustness rely neither from the NHSE nor nontrivial topology. Originally interpreted as enigmatic free-fermion eigenstates with negative probability~\cite{lee2022exceptional}, EB states arise from a totally new mechanism based on the geometric defectiveness of exceptional points,
where the eigenstates are not all linearly independent. While a $N\times N$-component Hermitian Hamiltonian is guaranteed full rank i.e. possess $N$ orthogonal eigenstates, a $N\times N$ non-Hermitian Hamiltonian can possess anywhere from one to $N$ linearly-independent eigenstates.

\begin{figure*}
		\includegraphics[width = \linewidth]{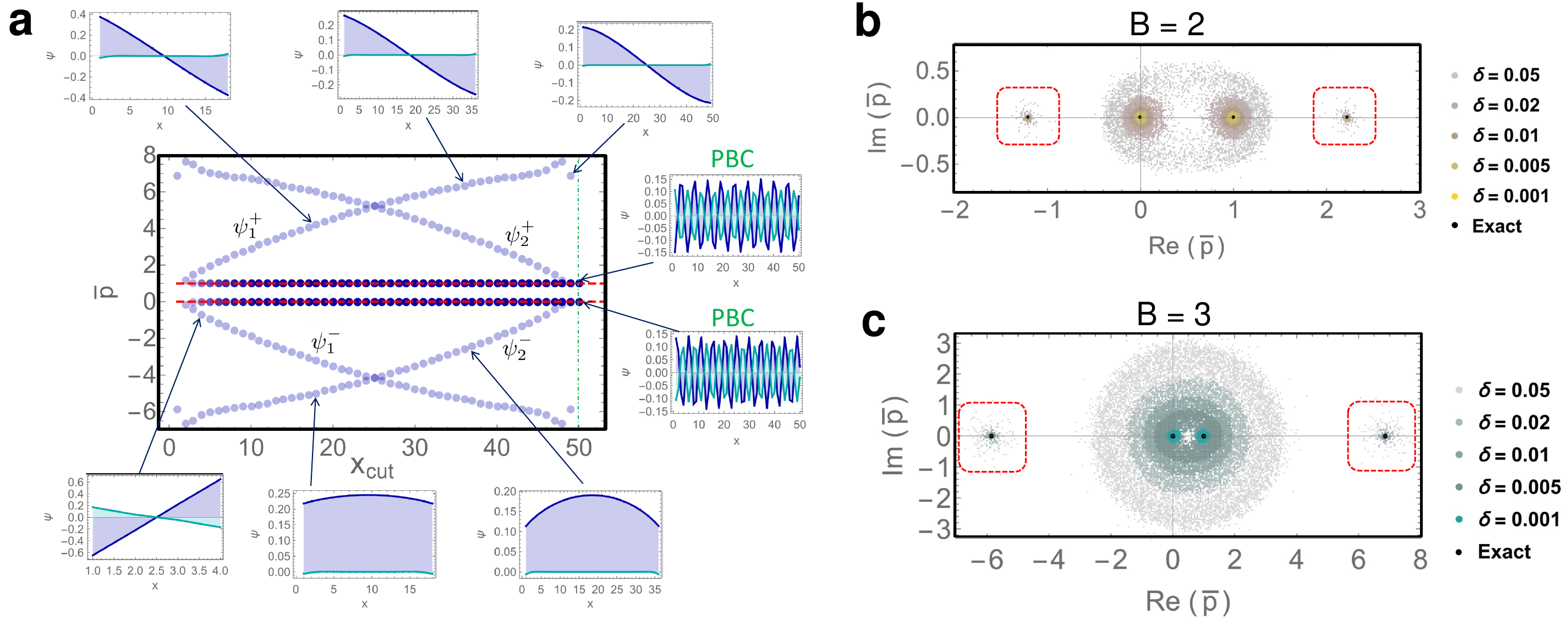}
		\caption{\textbf{Robustness of EB states and their spectral flow. } (a) Flow of occupation probability eigenvalues $\bar p$ for our canonical model $H_{n=2}$, $B=3$ as the un-truncated region $[x_L,x_R]=[1,x_{cut}]$ is varied. While most $\bar p$ cluster around $0$ and $1$, two distinct branches of EB states $\psi_1^\pm$ and $\psi_2^\pm$ exhibit distinct robust spectral flows away from them, with the $x_{cut}\leftrightarrow L-x_{cut}$ symmetry arising from the equivalent roles of the truncated vs. un-truncated regions. As $x_{cut}$ decreases, the EB eigenstates (insets) survive almost unchanged, testimony to the adiabatic continuity between EB states defined on very small and large $x_{cut}$ subsystems.
		(b,c) Distribution of the complex $\bar p$ eigenvalues from 50 instances of random geometric disorder in the $\bar P$ matrix elements with $x_{cut}=L-1$, with real and imaginary fractional perturbations taken uniformly from $[-\delta,\delta]$. Unlike the non-EB eigenvalues (central cloud), EB eigenvalues (red dashed) remain relatively unperturbed and well-gapped, even in (c) where the largest matrix elements scale like $L^{B-1}\sim \mathcal{O}(10^3)$. We have fixed $a_0=1$ and $L=50$ unit cells in all panels. 		
		}
    \label{fig:2}
\end{figure*}

We first define the EB states in a platform-independent manner. 
Unlike topological or skin states, EB states are not eigenstates of the Hamiltonian, but are instead special eigenstates of $\bar P$, the occupied Fermi sea projector [Fig.~\ref{fig:1}] 
restricted to a bounded region $[x_L,x_R]$~\cite{peschel2009reduced,pollmann2010entanglement}:
\begin{eqnarray}\label{barP}
\bar P &=& \sum_{x,x'\in [x_L,x_R]}|x'\rangle\langle c^\dagger_{x'}c_x\rangle\langle x|\notag\\
&=& \sum_{x,x'\in [x_L,x_R]}\left( \sum_{\mu \in \text{occ.}}\sum_k e^{ik(x'-x)}\mu_R(k)\mu^*_L(k)\right)|x'\rangle\langle x|.\notag\\
\end{eqnarray}
Specifically, EB states, which arise only when an exceptional point is present, are the eigenstates of $\bar P$ corresponding to eigenvalues far outside of $[0,1]$ (yellow in Fig.~\ref{fig:1} Bottom), the spectral range of an usual Hermitian projector. As numerically computed in Fig.~\ref{fig:2}a, they possess distinctive spatial profiles different from the exponential profiles of skin or topological states. Here $c^\dagger_{x}$ ($c_x$) creates (annihilates) a fermion at site $x$ and $\langle c^\dagger_{x'}c_x\rangle$ is its propagator within the occupied bands $\mu$, which are represented by right and left eigenstates defined by $H(k)\mu_R(k)=\epsilon_\mu(k)\mu_R(k)$ and $H^\dagger(k)\mu_L(k)=\epsilon^*_\mu(k)\mu_L(k)$. For a non-Hermitian Hamiltonian $H\neq H^\dagger$, the left and right eigenstates are not simply complex conjugates of each other; they are necessarily defined as such so that they form a biorthonormal basis satisfying $\mu'_L\cdot \mu^*_R=\delta_{\mu'\mu}$. Physically, this biorthogonal definition of the propagator $\langle c^\dagger_{x'}c_x\rangle$ preserves the probabilistic interpretation of quantum mechanics under non-Hermiticity~\cite{brody2013biorthogonal}. But paradoxically, it would also give rise to negative probability EB eigenstates as we shall shortly see.

We next review the notion of geometric defectiveness, such as to understand its fundamental role in ensuring EB robustness. Consider a gapless non-Hermitian Hamiltonian $H(k)$ with a $n$-fold degenerate eigenenergy $\epsilon_0$ (exceptional point) at $k=k_0$. If it were Hermitian, there would be $n$ linearly-independent eigenstates at $k_0$, all with eigenenergy $\epsilon_0$. 
However, since $H$ is non-Hermitian, the degenerate subspace at $k_0$ in general decomposes into a sum of Jordan blocks
, namely $H(k_0)=H_\text{non-deg}\oplus H_\text{deg}$ where $H_\text{non-deg}$ is gapped and $H_\text{deg}$ is degenerate and can be brought to the form $H_\text{JB}^1\oplus H_\text{JB}^2\oplus ...\oplus H_\text{JB}^j$ via a similarity transform. Each Jordan block $H_\text{JB}^i$ (with $i=1,...,j$) is $n_i\times n_i$ such that $\sum_{i=1}^j n_i=n$, and takes the form
\begin{equation}\label{JB}
H_\text{JB}^i=\left(\begin{matrix}
\epsilon_0 & 1 & 0 & \ldots & 0\\
0 & \epsilon_0 & 1 & \ldots & 0\\
0 & 0 & \epsilon_0 & \ddots & 0\\
\vdots & \vdots & \vdots & \ddots & 1\\
0 & 0 & 0 & 0 & \epsilon_0
\end{matrix}\right).
\end{equation}
Despite being $n_i\times n_i$, it possesses only a \emph{single} nonvanishing eigenstate $(1,0,0,\ldots,0)^T$, and is thus of rank $1$. 
As such, the $n\times n$ $H_\text{deg}$ and thus $H(k_0)$ has only $j\leq N$ independent eigenstates i.e. has geometric multiplicity $j$. When $j<N$, $H(k_0)$ is known as being geometrically \emph{defective}.


EB states precisely originate from the presence of special points $k_0$ known as exceptional points, where the Hamiltonian $H$ is geometrically defective. This causes the corresponding $\bar P$ to become singular, as explained below. From Eq.~\ref{JB}, it can be seen that within each Jordan block of the degenerate subspace, the left and right eigenstates take the forms $\mu_L=(0,...,1)^T$ and $\mu_R=(1,...,0)^T$, which are perfectly orthogonal. As such, the existence of nontrivial Jordan blocks i.e. geometric defectiveness makes the biorthogonal basis and thus the restricted Fermi sea projector $\bar P$ ill-defined. In particular, as $k_0$ is approached, $\bar P$ becomes more singular with increasingly divergent matrix elements (propagators) $\langle c^\dagger_{x'}c_x\rangle$. To concretely see this, we specialize to the simplest exceptional point ansatz with double degeneracy ($n=2$) in a single Jordan block, and consider perturbations around $k=k_0=0$:
\begin{align}
H_{n=2}(k)&=\left(\begin{matrix}
0 & a_0 + h(k)\\
h(k) & 0
\end{matrix}\right)\sim \left(\begin{matrix}
0 & a_0 \\
\frac1{2}k^{2B} & 0
\end{matrix}\right)
\label{Hn2}
\end{align}
with $h(k)=\frac1{2}(2(1-\cos k))^B$, which reduces to $H^1_\text{JB}$ with $n_1=2$ and $\epsilon_0=0$ at $k=0$ upon rescaling. Choosing the negative eigenenergy band to be occupied,  we obtain
\begin{align}
\langle c^\dagger_{x'}c_x\rangle_{n=2}&=\frac1{2}\sum_k e^{ik(x'-x)}\left(\begin{matrix}
1 & -\sqrt{\frac{a_0 + h(k)}{h(k)}}\\
-\sqrt{\frac{h(k)}{a_0 + h(k)}} & 1
\end{matrix}\right)\notag\\
&\sim \frac1{2}\sum_k e^{ik(x'-x)}\left(\begin{matrix}
1 & -k^{-B}\sqrt{2a_0}\\
-k^B/\sqrt{2a_0} & 1
\end{matrix}\right),
\label{cc2}
\end{align}
which is divergently asymmetric due to the $\sim k^{-B}$ contributions near $k=0$. Note that in the absence of $x_L,x_R$ boundaries, the momentum $k$ is diagonal and $\bar P$ would still have eigenvalues $\bar p=0$ and $1$ only, as expected for a Fermi sea projector. But with these $x_L,x_R$ boundary cutoffs, translation invariance is broken and the divergent $k^{-B}$ contributions from Eq.~\ref{cc2} dominate, yielding eigenvalues $\bar p$ that can be prominently outside of $[0,1]$.

These eigenvalues outside of $[0,1]$ are known as the EB eigenvalues (labeled as $\bar p_\text{EB}$), and their corresponding eigenstates dubbed EB states [Fig.~\ref{fig:1}]. For a lattice system with $L$ unit cells, the magnitude of $\bar p_\text{EB}$ scales like $L^{B-1}$, since the largest contributions to $\langle c^\dagger_{x'}c_x\rangle_{n=2}$ come from $k^{-B}$ at the momentum points $k=\sim \pm \pi/L$ nearest to $k=0$. This is more rigorously proven in the Methods. For every EB eigenvalue $\bar p_\text{EB}$, by construction $1-\bar p_\text{EB}$ is also an eigenvalue of $\bar P$. Arising due to the boundaries of $[x_L,x_R]$, the number of EB states do not scale with $L$: This is evident in Fig.~\ref{fig:2}a, which shows the $\bar P$ eigenvalues for $H_{n=2}$ (Eq.~\ref{Hn2}) with $B=3$, numerically plotted as a function of boundary position $[x_L,x_R]=[1,x_{cut}]$. Most eigenvalues (dark blue) remain within $[0,1]$, with the exception of isolated pairs of EB eigenvalues $\bar p_\text{EB}$ corresponding to EB eigenstates $\psi_1^\pm, \psi_2^\pm$ (pale blue). 

We next discuss the interpretation of the EB eigenvalues $\bar p_\text{EB}$. Since $\bar P$ projects onto the occupied bands within $[x_L,x_R]$, its eigenvalues $\bar p$ can be interpreted as the occupation probability of their corresponding eigenstates. Indeed, for free fermions, the departure of $\bar p$ from $0$ or $1$ represents the amount of quantum entanglement across the boundaries of $[x_L,x_R]$, 
as quantified by the entanglement entropy $S=-\sum_{\bar p}\left[ \bar p\ln \bar p + (1-\bar p)\ln(1-\bar p)\right]$~\cite{calabrese2012entanglement}, which is greatly contributed by values of $\bar p$ away from $0$ or $1$. For gapped systems, all eigenvalues $\bar p$ lie exponentially close to $0$ or $1$, leading to very little entanglement; for usual gapless systems without geometric defectiveness, $\bar p$ become distributed within $[0,1]$ [Fig.~\ref{fig:1} Bottom Left] with logarithmic spacings, leading to $S\sim \frac1{3}\log L$, where $L$ is the system size~\cite{gioev2006entanglement,guo2021entanglement}. However, at geometrically defective critical points, one or more pairs of eigenvalues $\bar p_\text{EB}\sim L^{B}$ lie prominently outside of $[0,1]$. Literally interpreted, the EB eigenvalue $1-\bar p_\text{EB}<0$ thus represents negative probability: this is possible because the band occupancy is characterized by a biorthogonal projector $\sum_{k,\mu\in \text{occ.}}|\mu_R(k)\rangle\langle \mu_L(k)|$ involving left/right eigenstates (bra/kets) that are generically different. Alternatively, we can also understand why we can have $1-\bar p_\text{EB}<0$, or equivalently $\bar p_\text{EB}>1$, through the divergently asymmetric 2-point function $\langle c^\dagger_{x'}c_x\rangle$. By having strongly asymmetric ``hoppings'' across the boundary of the cut region $[x_L,x_R]$, there can be a net gain of state amplitude into the region , resulting in $\bar p_\text{EB}>1$, somewhat reminiscent of Hawking radiation at the event horizon of a black hole~\cite{almheiri2020entropy,chen2020quantum}.


\subsection{Exceptional bound (EB) states as physical eigenstates}

Even though EB states possess interesting and unusual properties, to directly observe them, they have to be eigenstates of a physical observable, and not just the truncated projector $\bar P$. Hence, EB states have hitherto been considered physically elusive, since $\bar P$ contains extremely non-local matrix elements, and cannot be approximated by experimentally realistic Hamiltonians or network Laplacians.

What has not been realized, however, is that EB states can survive robustly even if $\bar P$ is extensively truncated, such that only a small proportion of the non-local matrix elements remain. Below, we establish this important fact, and furthermore show that EB states are relatively insensitive to system disorder compared to all other eigenstates. These two discoveries make it possible to observe EB states in rather small systems governed by experimentally realistic Hamiltonians or network Laplacians that mathematically simulate the heavily truncated $\bar P$, \emph{without} the need for constructing the 
geometrically defective parent systems.

\subsubsection{EB spectral flow and their robustness against spatial truncation}

EB spectra exhibits a novel type of spectral flow that gives EB states its unusual robustness. This is illustrated in Fig.~\ref{fig:2}a, which shows how the eigenvalues of $\bar P$ evolves as the position of its truncation cut $x_{cut}=x_R$ is varied from $1$ to $L=50$ (we fix $x_L=1$). 
For definiteness, we have used $H_{n=2}$ (Eq.~\ref{Hn2}) with exceptional point dispersion $\sim k^{2B}=k^6$, although other choices of $B$ generically give rise to qualitatively similar behavior [Fig.~\ref{fig:B} of Methods].

Saliently, while almost all eigenvalues $\bar p$ are infinitesimally closely locked to $0$ or $1$ (red dashed), there exist two pairs of eigensolutions ($\psi_1^\pm$ and $\psi^\pm_2$) that evolve away from $[0,1]$, crossing each other at $x_{cut}=L/2=25$ in a perfectly symmetrical manner. This ``EB spectral flow'' is reminiscent of the spectral flow of topological states as flux is threaded (see Refs.~\cite{niu1985quantized,qi2011generic,alexandradinata2011trace,yu2011equivalent,lee2015free}), even though there cannot be any topological pumping in such a critical system.

The existence of this spectral flow is very important for the practical physical demonstration of EB states, because it implies that an EB state which is well-defined at large truncated system sizes $x_{cut}$, when the lattice approximates a continuum, is adiabatically connected to a qualitatively similar EB state at very small truncated systems with $x_{cut}\sim \mathcal{O}(1)$. As such, bone-fide EB state phenomena can be demonstrated by truncating $\bar P$ to just a few unit cells, at which their conceptual origin becomes otherwise murky. Indeed, the $\psi_1^\pm$ EB state in the nearly-continuum case of $x_{cut}=49$ unit cells (top right) remains almost qualitatively unchanged as $x_{cut}$ gets smaller, keeping its characteristic sine-like profile (derived in the Methods) even down to $x_{cut}=4$ (bottom left). Its adiabatic continuity can be justified through the large spectral gap separating the EB states from the other states. The $\psi_2^\pm$ branch represents a different orthogonal eigensolution that is odd about $x_{cut}=L/2$. For $x_{cut}=L$ (green), we simply have PBCs and $\bar p$ reduces to exactly $0$ or $1$.

In the following, we sketch the origin of the robustness of this EB spectral flow; the interested reader is invited to refer to the Methods for a more complete proof. Unlike topological spectral flow, which is protected by quantized Wannier center transport~\cite{niu1985quantized,qi2011generic,alexandradinata2011trace,yu2011equivalent,lee2015free}, EB spectral flow is not protected by any topological invariant. Yet, the adiabatic continuity  and crisscrossing flows of the EB spectrum with $x_{cut}$ is equally inevitable due to the divergently long-ranged nature of the 2-point functions that form the matrix elements of $\bar P$. In the Methods, it was shown that the largest 2-point functions decay slower than a power-law (for $B>1$), and importantly also causes the EB states to exhibit approximately similar spatial profiles. Due to their very slow decay, the extent of entanglement, as captured by the departure of $\bar p_\text{EB}^2-\bar p_\text{EB}$ from zero, must therefore scale linearly with the size of the non-truncated region $x_{cut}$ (Eq.~\ref{lambda2}). This spectral flow behavior originates from the long-ranged 2-point functions, and cannot be destroyed unless the parent Hamiltonian is no longer geometrically defective. As such, the EB states from a heavily truncated $\bar P$ with very small number $x_{cut}$ of unit cells must still be adiabatically connected to EB states defined on a large lattice.


\subsubsection{Robustness of EB states against hopping disorder}

Besides surviving robustly even as the hopping model $\bar P$ is truncated down to very small sizes, EB states are also remarkably robust to random hopping disorder when $\bar P$ is taken as a physical Hamiltonian or Laplacian. A key feature in the hoppings of $\bar P$ is that it ranges over many orders of magnitude -- as derived in the Methods, the largest hoppings scale like $L^{B-1}$ (for $B>1$), while the smallest do not scale with $L$ at all. This dramatic contrast between strong and weak hoppings is what gives rise to the illusion of ``negative probability'' through a Hawking radiation-like argument; yet, it also brings about the very real prospect of experimental disorder in the few strongest hoppings drowning out all other subtle signatures.

Fortunately, EB states 
turn out to be far more resilent towards hopping disorder than non-EB states. To simulate component disorder in actual experiments, the disorder is introduced by multiplying each hopping by a random factor of $1+r_R+ir_I$, where $r_R$ and $r_I$ are real random variables uniformly selected from $[-\delta,\delta]$. In electrical circuit realizations, for instance, complex disorder can stem from stray capacitances or inductances that lead to random phase shifts. Spectral plots with only real or imaginary disorder can be found in the Supplement.

As shown in Figs.~\ref{fig:2}b and c for $H_{n=2}$ with $B=2$ and $B=3$ respectively, the $\bar P$ eigenvalues from 50 independent instances of geometric hopping disorder form dispersed spectral ``clouds''. At zero disorder, the eigenvalues $\bar p$ (black dots) are either infinitesimally close to $0$ or $1$ (non-EB states), or take on isolated values outside of $[0,1]$ (EB states). With nonzero fractional hopping disorder $\delta$, the non-EB eigenvalues quickly become very dispersed, purportedly because very small fractional uncertainties in the largest hoppings can be of large absolute magnitudes. Indeed, when the disorder fraction reaches 5$\%$ ($\delta=0.05$) for the $B=2$ case [Figs~\ref{fig:2}b], the $0$ and $1$ non-EB eigenvalues become so perturbed that their density clouds cannot be separated - alternatively, we say that with even 5$\%$ complex disorder, the notion of occupancy becomes ill-defined. For the $B=3$ case [Figs~\ref{fig:2}c] with more strongly divergent hoppings in $\bar P$, this threshold is lowered to about 0.5$\%$ ($\delta =0.005$), hardly a generous uncertainty allowance for realistic experimental components~\cite{lee2018topolectrical,lee2020imaging}.

Yet, from Figs.~\ref{fig:2}b and c, it is evident that the EB eigenvalues (red dashed) remain well-separated from non-EB eigenvalues even at $\delta=0.05$, when the notion of non-EB state occupancy is not even well-defined. The fractional spread in the EB eigenvalue ``cloud'' remains at approximately $\delta$ of the unperturbed $\bar p_\text{EB}$, seemingly impervious to the disorder from the largest hoppings. All in all, we have established the feasibility of realizing EB states in lattices with significant geometric disorder, even in regimes where it is hopeless to observe non-EB spectral signatures.

\subsection{Experimental observation of EB states}

In the parent system $H$ which contains an exceptional point, the truncated band projector $\bar P$ is not a physical observable and its EB eigenstates would be experimentally elusive. However, by constructing an alternative system described by a Hamiltonian or network Laplacian that is mathematically identical to $\bar P$, we can directly observe EB eigenstates as unique and isolated resonant peaks. 
Furthermore, instead of having a large number $L$ of unit cells as in the parent system, $\bar P$ only needs to have $x_{cut}\sim \mathcal{O}(1)$ unit cells to exhibit well-defined EB states.

\subsection{Description of experimental setup}

\begin{figure*}
    \centering
    \includegraphics[width = \linewidth]{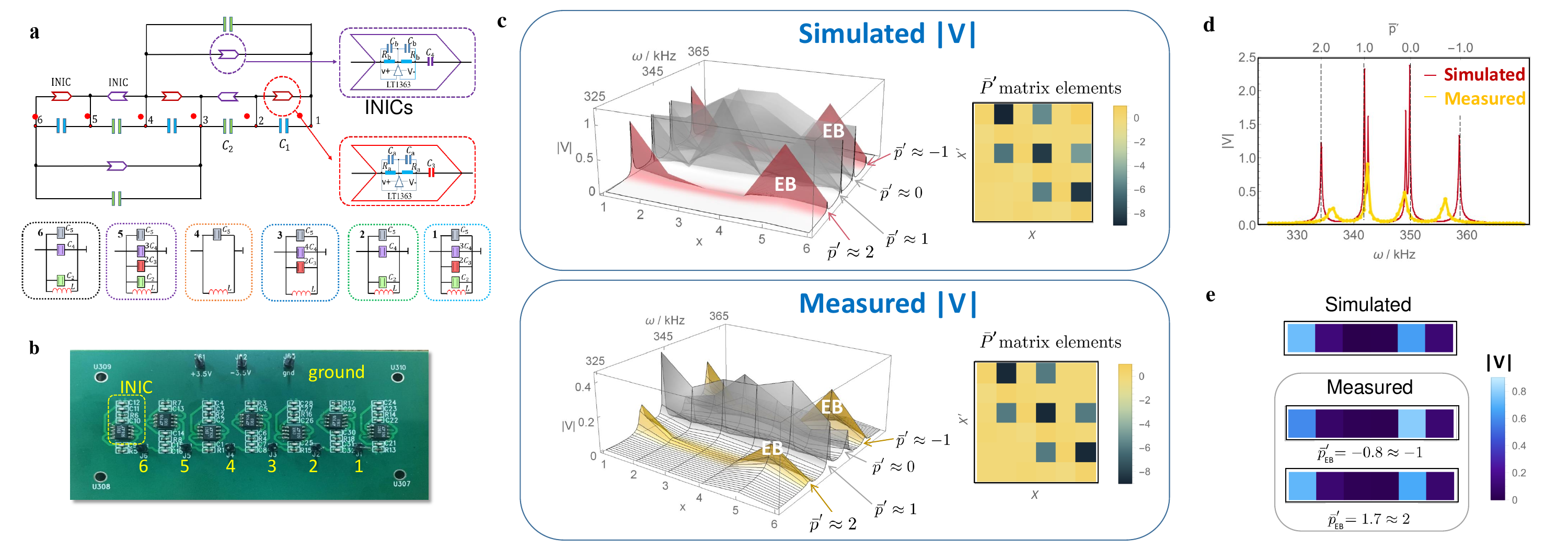}
    \caption{\textbf{Experimental measurement of EB states.} (a,b) Our LC circuit, which implements Eq.~\ref{pp1} via the Laplacian given by Eqs.~\ref{Jg} and \ref{Jc}, consists of 6 nodes connected and grounded (circled 1 to 6) by various combinations of capacitors and inductors. INICs containing non-Hermitian op-amps implement the requisite asymmetric couplings. (c) Simulated and measured voltage profiles as a function of AC frequency $\omega$, with resonances appearing at the frequencies corresponding to all four eigenvalue clusters $\bar p'\approx -1,0,1,2$ (Eq.~\ref{pp2}). However, only the voltage amplitude profiles at the EB eigenvalues $\bar p'\approx -1,2$ (colored ``EB'') resemble that of their corresponding eigenstates. Very good agreement (Right) exist between the theoretically simulated $\bar P'$ matrix elements and their reconstructed values from node-resolved impedance measurements. (d) Both the simulated (red) and measured (yellow) voltage amplitudes exhibit resonances at the four frequencies corresponding to $\bar P'$ eigenvalues, although the measured peaks are smoothed out by parasitic resistances. The peak positions agree to within 1\% (2-3 kHz). (e) Excellent quantitative agreement between the simulated EB state amplitude profile $|V|$, which is identical for both $\bar p'_\text{EB}\approx -1,2$, and their measured counterparts at slightly shifted $\bar p'_\text{EB}$ frequencies.
				}
    \label{fig:circuit}
\end{figure*}

We experimentally observed the EB states of our $H_{n=2}$ Hamiltonian by implementing its $\bar P$ via an electrical LC circuit network [Fig.~\ref{fig:circuit}a,b], with the non-Hermiticity introduced via operational amplifiers (op-amps) encapsulated in negative impedance converters with current inversion (INICs, see Methods). The steady state in electrical circuits is governed by Kirchhoff's law, which can be written in the matrix form $\bold I = J\bold V$, where $\bold I=(I_1,I_2,...)^T$ represents the input currents at nodes labeled $1,2,$ etc and $\bold V=(V_1,V_2,...)^T$ represent their corresponding electrical potentials. The matrix $J$, known as the circuit Laplacian, encodes the component connectivity of the circuit, and can be designed to accurately approximate the form of our matrix $\bar P$ as desired.

In our experiment, we chose $x_{cut}=3$ and design our circuit [Fig.~\ref{fig:circuit}a] such that the eigenvalue equation $\bar P \bold V = \bar p \bold V$ is approximately satisfied when strong topolectrical resonances~\cite{lee2018topolectrical} occur. In such scenarios, nonzero voltage potential profiles ($\bold V \neq \bold 0$) can exist even with negligible external source currents ($\bold I=0$), so the eigenvectors $\bold V$ are mapped into the kernel of the circuit Laplacian $J$, i.e. $J\bold V=\bold 0$. With details left to the Methods, the circuit thus designed and fabricated is illustrated in Figs.~\ref{fig:circuit}a,b, with 6 nodes (labeled $1$ to $6$) connected by various combinations of capacitors, inductors and op-amps. It gives rise to the following simplified eigenvalue equation 
\begin{equation}
\frac1{C_0}\left(\begin{matrix}
\bold C_{13} & \bold C_{24} & \bold 0 \\
\bold C_{24} & \bold C_{13} & \bold C_{24} \\
\bold 0 & \bold C_{24} & \bold C_{13}
\end{matrix}\right)\bold V =\bar P'\,\bold V=\bar p'\,\bold V
\label{pp1}
\end{equation}
where $\bar P'$ is a matrix that closely approximates to the exact $\bar P$ for parameters $a_0=1$, $B=7$, $L=4$ (see Supplement), as far as commercially available circuit components and their uncertainty tolerances allow. As elaborated in the Methods, its eigenvalue is related to the circuit parameters via
\begin{equation}
\bar p'=\frac1{\omega^2 LC_0}-\frac{C_1+2C_2+C_3+2C_4}{C_0},
\label{pp2}
\end{equation}
with $C_0=1$nF setting the capacitance scale.
We have $\bold C_{13} = C_5\,\mathbb{I}-C_1\sigma_x-i\sigma_yC_3$ and $\bold C_{24} = -C_2\sigma_x-i\sigma_yC_4$, where $\sigma_x, \sigma_y$ are the Pauli matrices and $C_1=4.55$nF, $C_2=2.87$nF, $C_3=4.3$nF, $C_4=3.04$nF and $C_5=0.50$nF are capacitors in the circuit. $L=10\mu$H are grounded inductors introduced such that $p'$ can be freely tuned by adjusting the AC frequency $\omega$ (Eq.~\ref{pp2}). To minimize component disorder, we have chosen to include only 5 unique capacitance parameters corresponding commercially available capacitors, truncating the other matrix elements to zero; with them, Eq.~\ref{pp1} gives EB eigenvalues $\bar p'_\text{EB}=1.987$ and $-0.987$, which are already extremely close to the exact EB eigenvalues $\bar p_\text{EB}=2.007$ and $-1.007$.

While the implementation of the positive LC components is straightforward, negative capacitors $-C_3$ and $-C_4$ are also present, as explicitly seen from their asymmetric appearances in the $i\sigma_y$ terms of $\bold C_{13}$ and $\bold C_{24}$ (Eq.~\ref{pp2}). The implementation of negative capacitive feedback requires more involved op-amp circuitry (see Methods) which are decidedly non-Hermitian, drawing power from external voltage sources. Note that this coupling asymmetry \emph{cannot} give rise to the NHSE because the PBC spectrum of the projector consists of isolated spectral points $0$ and $1$ by construction, and do not possess the spectral winding necessary for the NHSE~\cite{}.

\subsection{Experimental results}

To conclusively establish the observation of EB phenomena in our circuit, we performed the (i) measurement of EB spectra as topolectrical resonances; (ii) spatially mapping of the EB eigenstate from the electrical potential profile; and (iii) reconstruction of the circuit Laplacian $J$ corresponding to $\bar P'$. 

EB resonances in our circuit should be detectable via two-point impedance measurements between any two nodes. Given any two nodes $i$ and $j$, 
the impedance $Z_{ij}$ between them is given compactly by~\cite{lee2018topolectrical,helbig2020generalized}
\begin{eqnarray}
Z_{ij}
&=&\sum_\mu\frac{\left(\langle \psi_\mu^L(i)|-\langle \psi_\mu^L(j)|\right)\left(| \psi_\mu^R(i)\rangle-| \psi_\mu^R(j)\rangle\right)}{j_\mu}\qquad
\label{Zij}
\end{eqnarray}
where $j_\mu$ is the $\mu$-th eigenvalue of the circuit Laplacian $J$, with $|\psi_\mu^L\rangle,|\psi_\mu^R\rangle$ its corresponding left and right eigenstates. 
Since we have designed our circuit such that all $\bar P'$ eigenstates $\bold V$, particularly the EB eigenstates, satisfy $J\bold V=0$, they are precisely the same voltage profiles corresponding to $j_\mu=0$ i.e. divergent impedance $Z_{ij}$. In other words, when the AC frequency $\omega$ is adjusted such that $\bar p'$ satisfy the eigenvalue equation Eq.~\ref{pp1}, the two-point impedance between any two nodes should exhibit a resonance. Substituting the EB eigenvalues of $\bar p'_\text{EB}=1.987\approx 2$ and $-0.987\approx -1$ and non-EB eigenvalues $\bar p'=0.991,0.906,0.094$ and $0.009$ into Eq.~\ref{pp2}, we theoretical predict two resonant EB frequencies of 334kHz and 359kHz corresponding to $\bar p'\approx 2$ and $-1$ respectively, and four non-EB resonant frequencies 342.0kHz, 342.6kHz, 349.3kHz and 350 kHz. 

As shown in Fig.~\ref{fig:circuit}c, these predictions are indeed in very close agreement with the simulated and measured voltage profiles $\bold V=(V_1,...,V_6)^T$ at the nodes, when an AC current is driven through the circuit. For definiteness, we have chosen to connect node 2 and the ground with a AC signal of amplitude 2V, even though the $\bar p'$ eigenvalues corresponds to topolectrical resonances $j_\mu=0$ which should always cause a divergent voltage response, regardless of the input current configuration (Eq.~\ref{Zij}). In both the simulations and experimental measurements, pronounced resonances were detected at the said EB and non-EB frequencies $\omega$. This is more precisely plotted in Fig.~\ref{fig:circuit}d, which shows the qualitative agreement between the theoretically simulated resonance peaks (red) and the measured peaks (yellow), which are situated at $\omega=$336.6kHz ($\bar p'_\text{EB}=1.7$), 343.2kHz ($\bar p'=0.83$), 349.6kHz ($\bar p'=0.05$) and 356.9kHz ($\bar p'_\text{EB}=0.8$). Note that in terms of the absolute frequency $\omega$, the measured and theoretically computed peaks only differ by less than $1\%$, which is arguably excellent considering the effects of component uncertainties and parasitic resistances. While the latter have also broadened the measured peaks, the EB peaks remain well-resolved.

The EB state profiles can be directly read from the electrical potential profiles $\bold V$ at the resonant EB frequencies, since EB states are well-isolated and not susceptible to mixing. This is evidently shown in Fig.~\ref{fig:circuit}c, for both the simulated (red) and measured (yellow) voltage amplitude profiles at the EB frequencies $\bar p'=\bar p'_\text{EB}\approx 2$ or $-1$. Their excellent agreement can be more explicitly seen in Fig.~\ref{fig:circuit}e, which compares the normalized EB amplitude profiles (which are mathematically identical for $\bar p'_\text{EB}$ and $1-\bar p'_\text{EB}$) between experiment and theory. By contrast, the non-EB resonant profiles at $\bar p'\approx 0$ or $1$ are very different between the simulations and measurements, because both of these eigenvalues are doubly degenerate and thus possess no unique eigenstate. Note that we can only record the voltage amplitude $|V|$ profile, and not $V$, since its sign cannot be unambiguously determined from the oscilloscope.

To further confirm that we have experimentally realized the occupied band projector $\bar P'$ as a linearly transformed circuit Laplacian (see Eq.~\ref{JV} in the Methods), we performed node-to-ground impedance measurements to reconstruct the individual Laplacian matrix elements. By running a current $I_x=I_0\delta_{xi}$ from an arbitrary node $i$ to the ground and measuring the electrical voltage potentials $\bold V$ at all the nodes, one obtains the $i$-th column of the pseudoinverse $J^{-1}$ of the Laplacian viz. $\bold V = J^{-1}\bold I$. This allows $J^{-1}$ and hence $J$ to be measured, from which $\bar P'$ can be recovered via Eq.~\ref{JV}. Comparing the right panels of Fig.~\ref{fig:circuit}c, we see that most of the matrix elements of $\bar P'$ can indeed be accurately experimentally recovered.

\subsection{Discussion and outlook}
\label{sec:discussion}
Originally defined as enigmatic negative probability eigenstates of the free fermion entanglement Hamiltonian, which is not a physical operator, EB states lived in the domain of mathematical abstraction, being not observable in principle. Yet, by successfully realizing their parent bounded free fermion projector $\bar P$ in terms of the Laplacian of a \emph{classical} electrical circuit, we have managed to experimentally observe these elusive EB states as prominent resonance profiles. 

The feasibility of such a circuit implementation relied not just on the versatility of electrical circuit connectivity, but also to two hitherto undiscovered contributing avenues of EB robustness: (i) a novel spectral flow that links rigorously defined EB states in the continuum limit with their much simpler discrete counterparts in a much smaller system [Fig.~\ref{fig:2}b], and (ii) their far superior robustness against coupling disorder compared to non-EB states [Fig.~\ref{fig:2}c]. This heightened robustness, in particular, could find applications in the engineering of relative stable modes in systems that are otherwise intrinsically very sensitive. 

While reminiscent of the more familiar topological Wannier flow or non-Hermitian skin states in some ways, EB states are decidedly a distinct new phenomenon in classical settings. Despite exhibiting nontrivial EB ``polarization'' in $\bar p$ as the entanglement region is varied, this spectral flow is inherited from the long-ranged profile of its gapless 2-point functions [Fig.~\ref{fig:Bprofile}], and not gapped topological invariant. And, despite the presence of asymmetric couplings, EB systems are completely impervious to the NHSE because their PBC ($x_{cut}=L$) spectra consists of isolated points that cannot possibly collapse any further to accommodate the NHSE.

\section{Acknowledgements}
CHL and HYM acknowledge the National Research Foundation Singapore grant under its QEP2.0 programme (NRF2021-QEP2-02-P09), and the Ministry of education (MOE) Tier-II grant (Award No. MOE-T2EP50222-0003).

\section{Data Availability}
The data that support the findings of this study are available from the corresponding author upon reasonable request.

\section{Author Contributions}
DZ and TC performed the experiments and designed the circuits under the supervision of XZ. HM and YSA improved the results and participated in the manuscript writing.  CHL proposed the initial idea, performed the computations, wrote the manuscript and provided the overall supervision for the project. The manuscript reflects the contributions of all authors.

\section{Competing Interests}
The authors declare no competing interests.

\clearpage
\pagebreak

\small

\section*{Methods}
\label{sec:methods}


\subsection{Approximate analytic solution for EB eigenvalues and their spectral flow}

Our starting point is the simplest possible defective Hamiltonian ansatz~\cite{lee2022exceptional}
\begin{equation}
H_{n=2}(k)=\left(\begin{matrix}
0 & a_0 + h(k) \\
h(k) & 0
\end{matrix}\right)
\label{Hn2m}
\end{equation}
with $h(k)=\frac1{2}\left(2(1-\cos k)\right)^B\sim k^{2B}$, such that the matrix symbol of the projector $ P$ is given by Eq.~\ref{cc2} of the main text:
\begin{equation}
P(k)=\left(\begin{matrix}
\mathbb{I} & -U \\
-D & \mathbb{I}
\end{matrix}\right)
\label{P2m}
\end{equation}
where $U(k)=D(k)^{-1}=\sqrt{\frac{a_0+h(k)}{h(k)}}$. EB states arise when $P$, or equivalently $U=D^{-1}$, becomes real-space truncated into $\bar P$ (or equivalently $\bar U\neq \bar D^{-1}$). Essentially, this means that $\bar P$ is no longer a projector, and EB states encode the extent by which $\bar P^2\neq \bar P$. At this point, it can be observed that the operator
\begin{equation}
\Lambda=4(\bar P^2-\bar P)=\left(\begin{matrix}
\bar U\bar D-\mathbb{I} & 0\\
0 & \bar D\bar U-\mathbb{I}
\end{matrix}\right)
\label{Lambda}
\end{equation}
not only measures how much $\bar P$ fails to be a projector, but also allows its eigensolutions to be separately solved in a two smaller decoupled subspaces. EB eigenvalues $\bar p_\text{EB}$ directly correspond to special eigenvalues $\bar\lambda_\text{EB}=4\bar p_\text{EB}(\bar p_\text{EB}-1)$ of $\Lambda$ that are far from zero.

To proceed, we examine Eq.~\ref{Lambda} in real space, and borrow a few observations about the spatial decay behavior of $ U$ and $ D$ that would be shown in the next subsubsection:
\begin{itemize}
\item $U_x$ diverges with $L$, with $U_x\sim L^{B-1}$ for $B\geq 2$, and $U_x\sim \log L$ for $B=1$.
\item $U_x$ decays slowly in space, logarithmically for $B=1$, linearly for $B= 2$, and asymptotically quadratically for $B>2$.
\item $D_x$ does not diverge, and decays rapidly.
\end{itemize}
where $U_x$ and $D_x$ are the real-space matrices of $U$ and $D$, obtained via Fourier transformation. Putting these observations together, it is found that $\bar U\bar D -\mathbb{I}$ must be dominated by its first slowly-decaying column, since the subsequent columns are attenuated by the rapid decay of $D_x$ coefficients. Denoting an EB eigenstate as $\psi=(\psi_\uparrow,\psi_\downarrow)^T$, such that $( \bar U\bar D -\mathbb{I})|\psi_\uparrow\rangle=\bar\lambda_\text{EB} |\psi_\uparrow\rangle$ (and analogously for $|\psi_\downarrow\rangle$), the above considerations imply that $|\psi_\uparrow\rangle\approx U_x$ up to normalization, as verified in Fig.~\ref{fig:Bprofile}. With some algebra, it can also be shown that the EB eigenvalues occur in pairs $p_\text{EB}^\pm=\frac1{2}\left(1\pm \sqrt{1+\bar\lambda_\text{EB}}\right)$, where
\begin{equation}
\bar\lambda_\text{EB} \approx -\frac{\sum_{x,x',x''=0}D_{x}U_{-x'}D_{x'+x''}U_{-(x+x'')}}{\sum_{x=0}D_{x}U_{-x}},
\label{lambda}
\end{equation}
the summation taken over the un-truncated region $[x_L,x_R]$.

As an immediate corollary, for $B\geq 2$ such that $U_x\sim L^{B-1}$, we also asymptotically have $\bar p_\text{EB}\sim \sqrt{\bar\lambda_\text{EB}}\sim L^{(B-1)/2}$, since the numerator of $\bar\lambda_\text{EB}$ is quadratic in the $U_x$, while the denominator is linear in them.

\subsubsection{Asymptotic behavior of the two-point function}

\begin{figure}
    \centering
    \subfloat[]{\includegraphics[width = .49\linewidth]{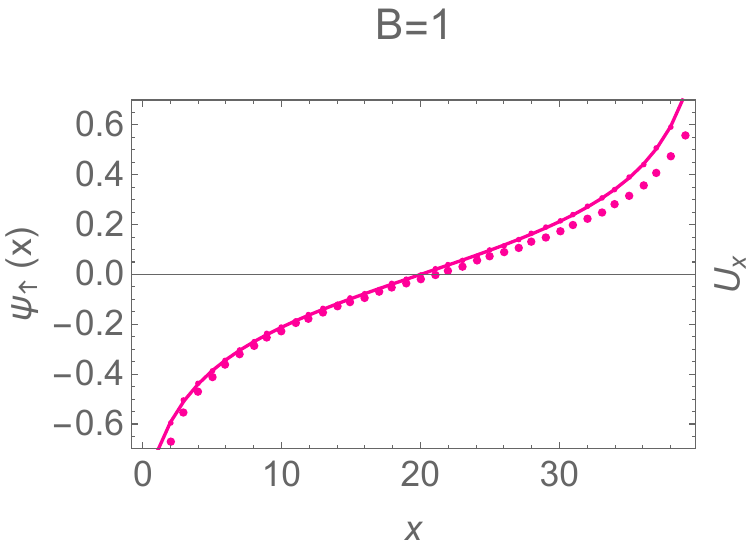}}
		\subfloat[]{\includegraphics[width = .49\linewidth]{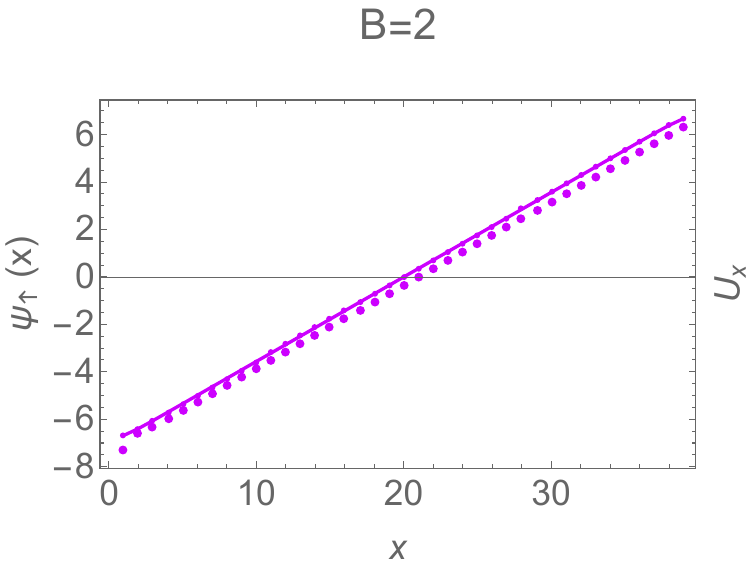}}\\
		\subfloat[]{\includegraphics[width = .49\linewidth]{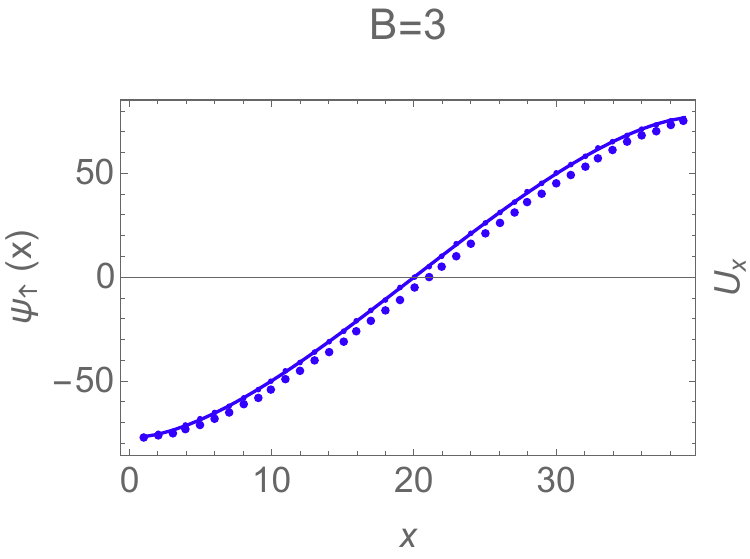}}
		\subfloat[]{\includegraphics[width = .49\linewidth]{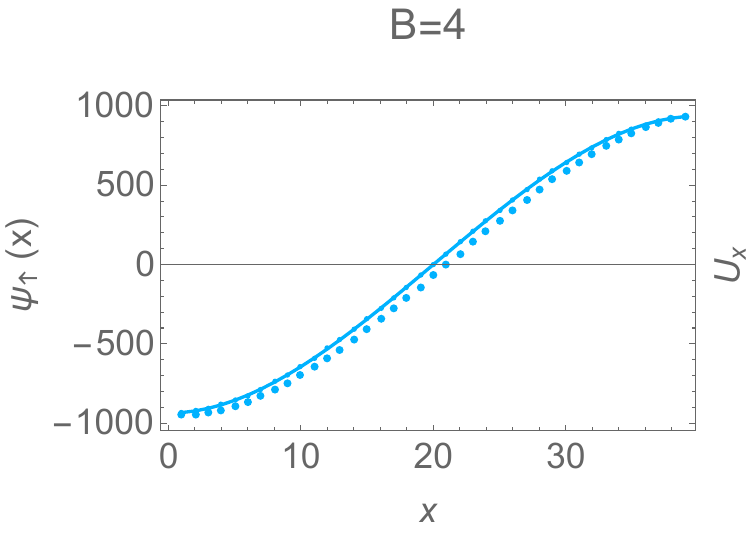}}
		\caption{\textbf{The similar spatial profiles of 2-point function $U_x$ and the leading EB state.} Shown in (a)-(d) are the spatial profiles of the 2-point function $U_x=-2\langle c^\dagger_{x,\uparrow}c_{0,\downarrow}\rangle$ (dotted curves), juxtaposed against the largest-eigenvalue EB eigenstates $\psi_\uparrow(x)$ of $\bar P$ (solid curves). Evidently, $\psi_\uparrow(x)\propto U_x$, as expected, since $U_x$ forms a dominating column in the matrix $\bar P$. For both $B=3,4$ (and in fact all higher $B$), the profiles are quadratic at small or large $x$.
		}
    \label{fig:Bprofile}
\end{figure}

Here, we provide some details on the asymptotic scaling properties of the two-point function, which are the matrix elements of $\bar P$. From Eq.~\ref{P2m}, the two-point function between the $\uparrow,\downarrow$ sublattices are given asymmetrically by Fourier coefficients $-U_x$ and $-D_x$. We have
\begin{eqnarray}
U_x &=& -2\langle c^\dagger_{x,\uparrow}c_{0,\downarrow}\rangle \notag\\
&\approx & \sqrt{a_0}\int_{\pi/L}^L\frac{e^{ikx}+e^{-ikx}dk}{(2\sqrt{1-\cos k})^B}
\label{int}
\end{eqnarray}
which simplifies to
\begin{eqnarray}
U_x|_{B=1}&\sim & 2\sqrt{a_0}\log \frac{L}{\pi x}-\gamma,\\
U_x|_{B=2}&\sim & 2\sqrt{a_0}\left( \frac{L}{x}-\pi x\right)\\
U_x|_{B>2}&\sim & 4\sqrt{a_0}\left(\frac{L}{\pi}\right)^{B-1}\left( 1-\frac{\pi^2x^2}{2L^2}\right).
\end{eqnarray}
For $B=1$, $\gamma=\int_1^\infty (\lfloor k \rfloor^{-1}-k^{-1})dk\approx 0.577$ is the Euler-Mascheroni constant.
For $B\geq 2$, we see from Eq.~\ref{int} that lattice momentum regularization sets the dominant contribution to be $U_x\sim L^{B-1}$, which diverges with $L$ (this result has nothing to do with the cut position $x_{cut}$). By taking the next order approximation with $x<L$, we obtain asymptotically quadratic scaling with $x$ for all $B>2$, except the $B=2$ case exhibiting a special linear scaling. Beyond $x>L/2$, the scaling behavior takes on its antisymmetric ``dual'' value under $x\rightarrow L-x$, as shown in Fig.~\ref{fig:Bprofile}, which confirms that the above spatial behavior of $U_x$ also approximately applies to that of their corresponding EB states.
The Fourier coefficients $D_x$, however, do not diverge since there is no divergent singularity in $D(k)=U(k)^{-1}$. As such, they remain small and decay exponentially with $x$.


\subsubsection{Spectral flow of $\bar p_\text{EB}$ with respect to $x_{cut}$}

In principle, Eq.~\ref{lambda} already allows one to make a reasonable accurate prediction of the value of $\bar\lambda_{EB}$ and hence $\bar p_\text{EB}$. However, given the universality of the decay behavior of $U_x$ and $D_x$, not just for our illustrative model but also for generic exceptional point dispersions, further conclusions are possible.

Since the other 2-point function $D_x$ exhibits rapid decay, the dominant contributions in Eq.~\ref{lambda} must contain terms with minimal range of $D_x$, i.e. $x=0$ and $x'=-x''$ for the numerator and $x=0$ for the denominator. In particular, for relatively small $x_{cut}/L$, that allows us to further approximate Eq.~\ref{lambda} by
\begin{eqnarray}
\bar\lambda_\text{EB} &\sim& -\frac{D_0}{U_0}\sum_{x''}^{x_{cut}}U_{x''}U_{-x''}+\mathcal{O}\left(e^{-2/B}\right)\notag\\
&\approx & U_0D_0\,x_{cut} \notag\\
&\sim& \left(\frac{L}{\pi}\right)^{B-1}\times x_{cut}
\label{lambda2}
\end{eqnarray}
where we have assumed that $B>1$. In the first line, we have only kept the $D_{x=0}$ term, and neglected all subleading terms. In the second line, we note that $U_x$ changes slowly with $x''$, especially at small $x''$, and the sum should thus approximately scale like the number of terms $x_{cut}$. In practice, this constant summand approximation works well even for $x_{cut}$ comparable to $L$.

In summary, we expect to have an eigensolution $\bar\lambda_\text{EB}=4\bar p_\text{EB}(1-\bar p_\text{EB})$ which scales approximately linearly with the size of the un-truncated region $x_{cut}$, which is defined by $[x_L,x_R]=[1,x_{cut}]$. And adiabatic continuity is expected due to the ``smooth'' linear dependence on $x_{cut}$. By construction, this argument pertains to only \emph{one} pair of EB eigenvalues: there can be other EB spectral branches that behave very differently.

\begin{figure*}[!t]
    \centering
    \subfloat[]{\includegraphics[width = .19\linewidth]{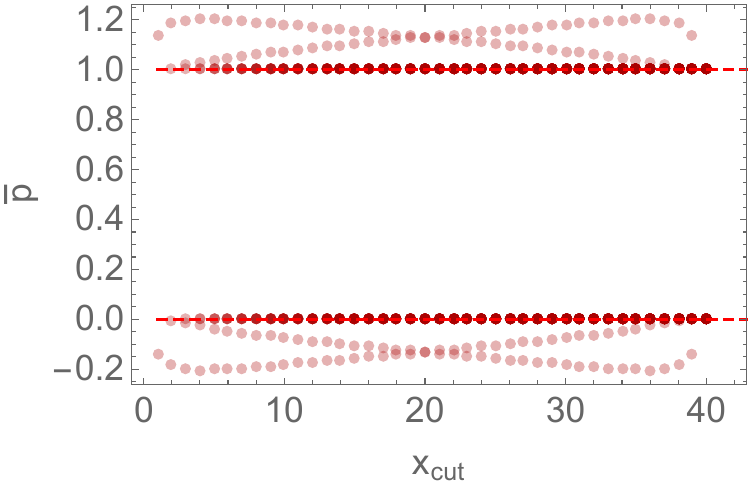}}
		\subfloat[]{\includegraphics[width = .19\linewidth]{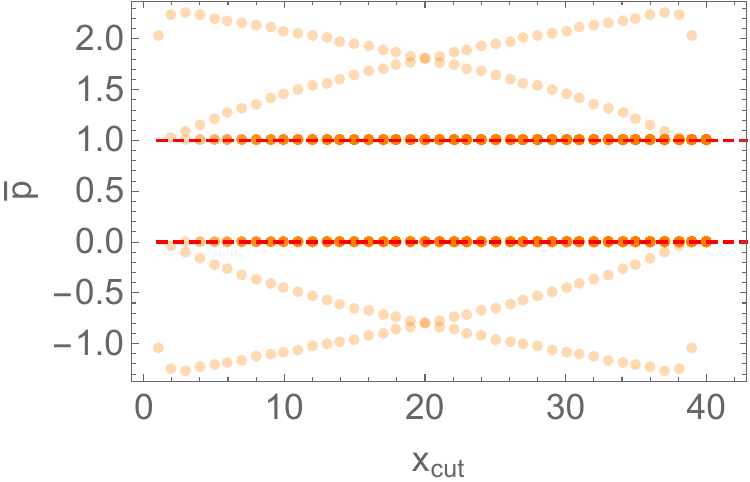}}
		\subfloat[]{\includegraphics[width = .19\linewidth]{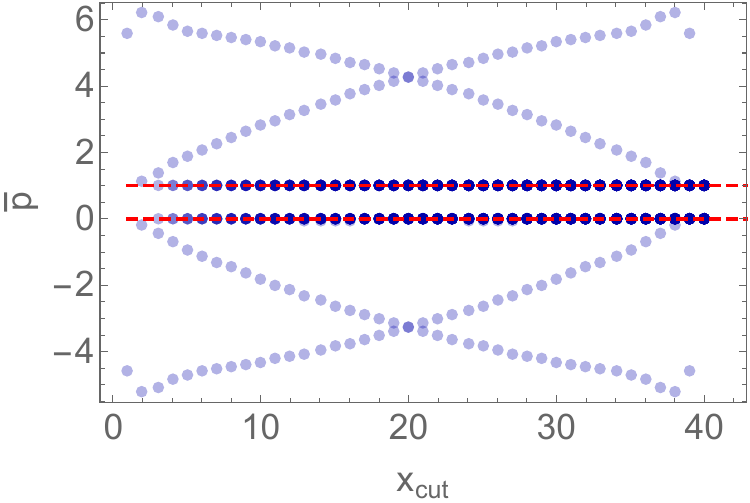}}
		\subfloat[]{\includegraphics[width = .19\linewidth]{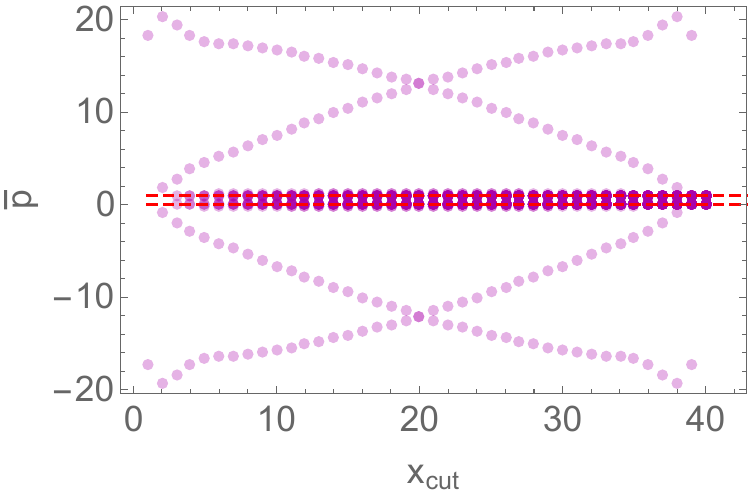}}
		\subfloat[]{\includegraphics[width = .19\linewidth]{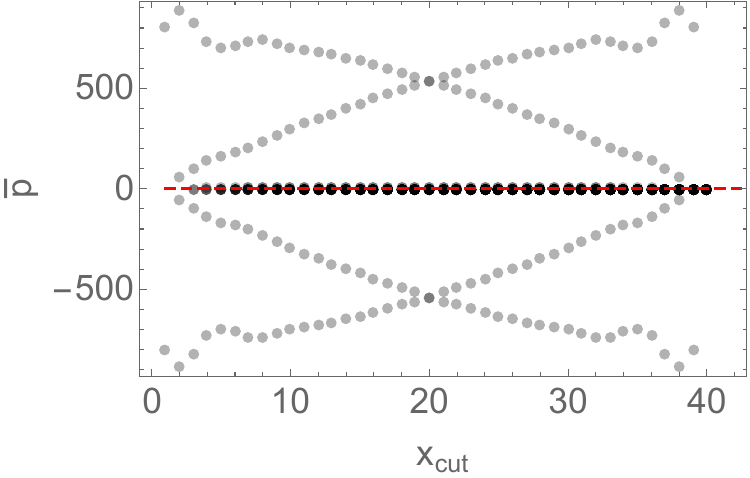}}
    \caption{\textbf{Effect of changing the order $2B$ of the exceptional point.} The $\bar p$ spectrum of our canonical model $H_{n=2}$ as the truncation position $x_{cut}$ is varied. From (a) to (e), we have $B=1,2,3,4,7$ respectively, all computed with $L=40$ and $a_0=1$. The values of $\bar p_\text{EB}$ increases exponentially with $B$, as predicted by Eq.~\ref{lambda2}, even though they qualitatively unchanged. Saliently, the EB branches increase in a slightly concave manner for all cases (together with their mirror images), which is in good numerical agreement with the theoretical prediction $\bar p_\text{EB}\propto 1\pm \sqrt{1+\bar\lambda_\text{EB}}\sim \left(\frac{L}{\pi}\right)^{B/2} \sqrt{x_{cut}}$ from Eq.~\ref{lambda2}, with $L/\pi=12.73$.
		}
    \label{fig:B}
\end{figure*}

To complete the justification of the EB spectral flow, we note that since swapping the truncated and un-truncated regions should not physically affect their entanglement, an linearly increasing $\bar \lambda_{EB}$  (i.e. square-root increasing $\bar p_\text{EB}$) solution must be accompanied by another identical but decreasing solution with $x_{cut}$ and $L-x_{cut}$ swapped. Indeed, this is what is numerically observed, as in Figs.~\ref{fig:2}, \ref{fig:B} and plots within the Supplement.

Interestingly, because of this ``reflection'' symmetry, for the smallest possible truncation with $x_{cut}=1$ unit cell, the two sole eigenvalues must be equivalent to the largest $\bar p_\text{EB}$ at $x_{cut}=L-1$. This is a vivid example of how EB behavior in the largest un-truncated systems is ``dual'' to the smallest one unit-cell $\bar P$ system.

\subsection{Details on the circuit observation of EB states}

\subsubsection{Truncated projector from circuit Laplacian}
\noindent In our experiment, we chose the parameters $x_{cut}=3$, $L=4$, $a_0=14$ and $B=7$ such as to obtain a $\bar P$ with EB eigenvalues that are approximately $2$ and $-1$, in addition to two pairs of non-EB eigenvalues close to $0$ and $1$. Explicitly, $\bar P=$
\begin{equation}\left(
\begin{array}{cccccc}
 0.5 & -8.8529 & 0 & -5.9055 & 0 & 0 \\
 -0.2566 & 0.5 & 0.1712 & 0 & 0 & 0 \\
 0 & -5.9055 & 0.5 & -8.8529 & 0 & -5.9055 \\
 0.1712 & 0 & -0.2566 & 0.5 & 0.1712 & 0 \\
 0 & 0 & 0 & -5.9055 & 0.5 & -8.8529 \\
 0 & 0 & 0.1712 & 0 & -0.2566 & 0.5 \\
\end{array}
\right)
\label{barPccc}
\end{equation}
with EB eigenvalues $\bar p_\text{EB}=2.0073,-1.0073$ and doubly degenerate non-EB eigenvalues $\bar p=0,1$. Note that with the chosen parameters, some off-diagonal matrix elements are also negligibly small. Next, we attempt to design a circuit with not more than four unique capacitance coupling values, such that its circuit Laplacian $J$ has all these $\bar P$ eigenvectors in its kernel to an excellent approximation. By considering commericially available capacitor values, our designed Laplacian is $J=J_g + J_c$, comprising a grounded part
\begin{widetext}
\begin{equation}J_g=\left[\frac1{i\omega L}+i\omega C_5\right]\mathbb{I}_{6\times 6}+i\omega\left(
\begin{array}{cccccc}
C_2+C_4 & 0 & 0 & 0 & 0 & 0 \\
0 & C_2+2C_3+3C_4 & 0 & 0 & 0 & 0 \\
0 & 0 & 0 & 0 & 0 & 0 \\
0 & 0 & 0 & 3C_3+4C_4 & 0 & 0 \\
0 & 0 & 0 & 0 & C_2+C_4 & 0 \\
0 & 0 & 0 & 0 & 0 & C_2+2C_3+3C_4
\end{array}
\right)
\label{Jg}
\end{equation}
and a inter-node coupling part exclusively involving capacitors:
\begin{equation}J_c=\left(
\begin{array}{cccccc}
C_1+C_2+C_3+C_4 & -C_3\!-\!C_1 & 0 & -C_4\!-\!C_2 & 0 & 0 \\
C_3\!-\!C_1 & C_1+C_2\!-\!C_3\!-\!C_4 & C_4\!-\!C_2 & 0 & 0 & 0 \\
0 & -C_2\!-\!C_4 & C_1+2C_2+C_3+2C_4 & -C_1\!-\!C_3 & 0 & -C_2\!-\!C_4 \\
C_4\!-\!C_2 & 0 & C_3\!-\!C_1 & C_1+2C_2\!-\!C_3\!-\!2C_4 & C_4\!-\!C_2 & 0 \\
0 & 0 & 0 & -C_2\!-\!C_4 & C_1+C_2+C_3+C_4 & -C_1\!-\!C_3 \\
0 & 0 & C_4\!-\!C_2 & 0 & C_3\!-\!C_1 & C_1+C_2\!-\!C_3\!-\!C_4
\end{array}
\right)
\label{Jc}
\end{equation}
where $C_1=4.55$nF, $C_2=2.87$nF, $C_3=4.3$nF, $C_4=3.04$nF and $C_5=0.50$nF. The inductor $L=10\mu$H introduces a relative factor of $\omega^{-2}$ that allows the AC frequency $\omega$ to be effectively used as a tuning knob for shifting the eigenvalues.

For states $\bold V$ in the kernel of $J$, $J\bold V =\bold 0$ is designed to recover the eigenvalue equation for $\bar P$. Due to limitations on the commercially available capacitors, this would only be approximate: we call the approximately obtained truncated projector $\bar P'$. From the above, $(J_c+J_g)\bold V =\bold 0$ can be re-arranged as
\begin{equation}
\frac1{C_0}\left[\frac1{\omega^2 L}-(C_1+2C_2+C_3+2C_4)\right]\bold V =-\left(\begin{matrix}
-C_5 & C_1+C_3 & 0 & C_2+C_4 & 0 & 0 \\
C_1-C_3 & -C_5 & C_2-C_4 & 0 & 0 & 0 \\
0 & C_2+C_4 & -C_5 & C_1+C_3 & 0 & C_2+C_4 \\
C_2-C_4 & 0 & C_1-C_3 & -C_5 & C_2-C_4 & 0 \\
0 & 0 & 0 & C_2+C_4 & -C_5 & C_1+C_3 \\
0 & 0 & C_2-C_4 & 0 & C_1-C_3 & -C_5
\end{matrix}\right)\bold V
\label{JV}
\end{equation}
or simply $\bar p'\bold V = \bar P'\bold V$, where $\bar p'=\frac1{C_0}\left[\frac1{\omega^2 L}-(C_1+2C_2+C_3+2C_4)\right]$ and $\bar P'$ is the matrix on the RHS (Eqs.~\ref{pp1} and \ref{pp2}). $C_0=1$nF normalizes the capacitances such that the non-EB eigenvalues are very close to $0$ and $1$. Numerically,
\begin{equation}
\bar P'=\left(
\begin{array}{cccccc}
 0.5 & -8.85 & 0 & -5.91 & 0 & 0 \\
 -0.25 & 0.5 & 0.17 & 0 & 0 & 0 \\
 0 & -5.91 & 0.5 & -8.85 & 0 & -5.91 \\
 0.17 & 0 & -0.25 & 0.5 & 0.17 & 0 \\
 0 & 0 & 0 & -5.91 & 0.5 & -8.85 \\
 0 & 0 & 0.17 & 0 & -0.25 & 0.5 \\
\end{array}
\right)
\end{equation}
which differs very little from the exact $\bar P$ given by Eq.~\ref{barPccc} above.
\end{widetext}

\subsubsection{Experimental design and procedure}

The experimental circuit is designed using the PADS circuit design software, which assists in selecting the PCB configuration, stackup layout, internal layers and grounding. To prevent unintended interference between different PCB layers, a grounding layer is placed in the gap between any two layers. Moreover, all PCB traces are designed to have a relatively large width (0.5mm) to reduce the parasitic inductance, and the spacings between electronic components are also made sufficient large (1.0mm) to avoid spurious inductive coupling.  A relay is used to simultaneously switch on or off the input voltages across all nodes when required. The op-amps used are of the model LT1363, and are powered through the DC power supply UNI-T UTP1306S. 

To ensure accurate experimental implementation of our theoretically designed circuit Laplacian (Eqs.~\ref{Jg}, \ref{Jc}), the component uncertainties have to be minimized and their stability optimized. Capacitors of requisite capacitances $C_1=4.55$nF, $C_2=2.87$nF, $C_3=4.3$nF, $C_4=3.04$nF and $C_5=0.50$nF with 0.5\% error tolerance are obtained by selecting from a large sample of commercially available 3.0nF and 4.7nF capacitors with 10\% error tolerance; C0G ceramic capacitors are chosen for their stability. 

The inductors contribute a major source of dissipation that can smooth out the desired impedance resonance peaks. To minimize that loss, we select only the inductors with the highest Q values at the operational frequency with impedances of no more than 200m$\Omega$, as measured  through a WK6500B impedance analyzer.

To measure the resonances and hence map out the EB state profiles, an input AC signal is input at an arbitrary node, and a digital storage oscilloscope (Agilent Technologies Infiniivision DSO7104B) measures the voltage signal $\bold V$ at each node. The input signal, which is chosen to be at node 2, is generated by an arbitrary waveform generator (ROGOL DG1022Z) with amplitude 2V, with frequency swept from 310kHz to 400kHz to accommodate all expected resonances ($\bar P'$ eigenvalues).  

To map out the matrix elements of the circuit Laplacian $J$, the same arbitrary waveform generator is used to successively input a signal with amplitude 2V and frequency 350kHz into each node, one node at a time. For a chosen input node $i$, the current is $I_j\propto \delta_{ij}$, and the relation $\bold V=J^{-1}\bold I$ allows the $i$-th column of $J^{-1}$ to be read by measuring the resultant voltages $\bold V$ on all the nodes. By iterating over $i$, $J^{-1}$ can be reconstructed, thereby yielding the Laplacian and its corresponding $\bar P'$ through matrix inversion.

\subsubsection{Implementation of asymmetric capacitive couplings through INICs}

\begin{figure}
    \centering
    \includegraphics[width = .8\linewidth]{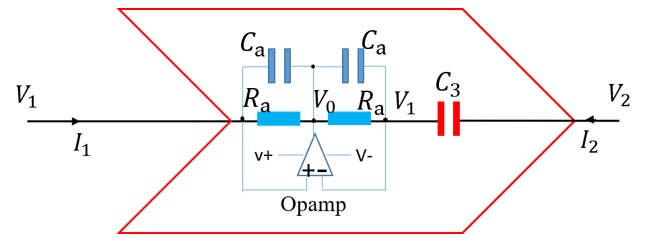}
    \caption{\textbf{Anatomy of an INIC} An INIC component in our circuit behaves like an usual capacitor $C_3$ from the right ($V_2$), but behaves like a negative capacitor $-C_3$ from the left ($V_1$). It consists of an op-amp connected to two identical feedback loops, which implement the non-reciprocal non-Hermiticity by allowing for the non-conservation of current.  
		}
    \label{fig:INIC}
\end{figure}

Asymmetric terms in the circuit Laplacian are implemented via negative impedance converter through current inversion (INICs), which simulates negative capacitance in one direction, while maintaining positive capacitance in the other direction. As shown in Fig.~\ref{fig:INIC}, the main ingredient of an INIC is an operational amplifier (op-amp); for experimental implementation we used the model LT1363. In the negative feedback configuration of the op-amp, the potentials at its positive and negative terminals are almost equal at $V_1$, with negligible current flowing into them. As such, we can write down the currents $I_1$ and $I_2$ as
\begin{eqnarray}
I_1&=&\frac1{Z_+}(V_1-V_0)\notag\\
I_2&=&\frac1{Z_-}(V_1-V_0)=i\omega C_3(V_2-V_1)
\end{eqnarray}
where $Z_+,Z_-$ are the total impedances of the positive and negative feedback loops. In general, $Z_+$ and $Z_-$ can take arbitrary values, depending on the components in these feedback loops, but for our application, it suffices to set them to be identical. Letting the feedback loops to consist of a parallel configuration of a resistor $R_a$ and a capacitor $C_a$, such as to avoid resonances at high frequencies, we have $Z_+=Z_-=(1/R_a+1/(i\omega C_a))^{-1}$. By writing the current-voltage relations in matrix form, we have
\begin{eqnarray}
\left(\begin{matrix}
I_1 \\ I_2 
\end{matrix}\right) &=& i\omega C_3
\left(\begin{matrix}
-Z_-/Z_+ & Z_-/Z_+ \\ -1 & 1  
\end{matrix}\right) 
\left(\begin{matrix}
V_1 \\ V_2 
\end{matrix}\right) \notag\\
 &=& i\omega C_3
\left(\begin{matrix}
-1 & 1 \\ -1 & 1  
\end{matrix}\right) 
\left(\begin{matrix}
V_1 \\ V_2 
\end{matrix}\right),
\end{eqnarray}
which manifestly contains an asymmetric Laplacian matrix. In particular, $I_1=I_2$ instead of the usual case of $I_1=-I_2$, where the current flow is conserved. All in all, the net effect is that this INIC behaves like an usual capacitor $C_3$ from the right, but behaves like a negative capacitor $-C_3$ from the left. Combined with other circuit elements, the INIC allows for the possibility of arbitrarily asymmetric circuit components. 

\bibliography{references}

\begin{thebibliography}{52}%
\makeatletter
\providecommand \@ifxundefined [1]{%
 \@ifx{#1\undefined}
}%
\providecommand \@ifnum [1]{%
 \ifnum #1\expandafter \@firstoftwo
 \else \expandafter \@secondoftwo
 \fi
}%
\providecommand \@ifx [1]{%
 \ifx #1\expandafter \@firstoftwo
 \else \expandafter \@secondoftwo
 \fi
}%
\providecommand \natexlab [1]{#1}%
\providecommand \enquote  [1]{``#1''}%
\providecommand \bibnamefont  [1]{#1}%
\providecommand \bibfnamefont [1]{#1}%
\providecommand \citenamefont [1]{#1}%
\providecommand \href@noop [0]{\@secondoftwo}%
\providecommand \href [0]{\begingroup \@sanitize@url \@href}%
\providecommand \@href[1]{\@@startlink{#1}\@@href}%
\providecommand \@@href[1]{\endgroup#1\@@endlink}%
\providecommand \@sanitize@url [0]{\catcode `\\12\catcode `\$12\catcode
  `\&12\catcode `\#12\catcode `\^12\catcode `\_12\catcode `\%12\relax}%
\providecommand \@@startlink[1]{}%
\providecommand \@@endlink[0]{}%
\providecommand \url  [0]{\begingroup\@sanitize@url \@url }%
\providecommand \@url [1]{\endgroup\@href {#1}{\urlprefix }}%
\providecommand \urlprefix  [0]{URL }%
\providecommand \Eprint [0]{\href }%
\providecommand \doibase [0]{https://doi.org/}%
\providecommand \selectlanguage [0]{\@gobble}%
\providecommand \bibinfo  [0]{\@secondoftwo}%
\providecommand \bibfield  [0]{\@secondoftwo}%
\providecommand \translation [1]{[#1]}%
\providecommand \BibitemOpen [0]{}%
\providecommand \bibitemStop [0]{}%
\providecommand \bibitemNoStop [0]{.\EOS\space}%
\providecommand \EOS [0]{\spacefactor3000\relax}%
\providecommand \BibitemShut  [1]{\csname bibitem#1\endcsname}%
\let\auto@bib@innerbib\@empty
\bibitem [{\citenamefont {Lee}(2022)}]{lee2022exceptional}%
  \BibitemOpen
  \bibfield  {author} {\bibinfo {author} {\bibfnamefont {C.~H.}\ \bibnamefont
  {Lee}},\ }\bibfield  {title} {\bibinfo {title} {Exceptional bound states and
  negative entanglement entropy},\ }\href@noop {} {\bibfield  {journal}
  {\bibinfo  {journal} {Physical Review Letters}\ }\textbf {\bibinfo {volume}
  {128}},\ \bibinfo {pages} {010402} (\bibinfo {year} {2022})}\BibitemShut
  {NoStop}%
\bibitem [{\citenamefont {Chang}\ \emph {et~al.}(2020)\citenamefont {Chang},
  \citenamefont {You}, \citenamefont {Wen},\ and\ \citenamefont
  {Ryu}}]{chang2020entanglement}%
  \BibitemOpen
  \bibfield  {author} {\bibinfo {author} {\bibfnamefont {P.-Y.}\ \bibnamefont
  {Chang}}, \bibinfo {author} {\bibfnamefont {J.-S.}\ \bibnamefont {You}},
  \bibinfo {author} {\bibfnamefont {X.}~\bibnamefont {Wen}},\ and\ \bibinfo
  {author} {\bibfnamefont {S.}~\bibnamefont {Ryu}},\ }\bibfield  {title}
  {\bibinfo {title} {Entanglement spectrum and entropy in topological
  non-hermitian systems and nonunitary conformal field theory},\ }\href@noop {}
  {\bibfield  {journal} {\bibinfo  {journal} {Physical Review Research}\
  }\textbf {\bibinfo {volume} {2}},\ \bibinfo {pages} {033069} (\bibinfo {year}
  {2020})}\BibitemShut {NoStop}%
\bibitem [{\citenamefont {Dembowski}\ \emph {et~al.}(2004)\citenamefont
  {Dembowski}, \citenamefont {Dietz}, \citenamefont {Gr{\"a}f}, \citenamefont
  {Harney}, \citenamefont {Heine}, \citenamefont {Heiss},\ and\ \citenamefont
  {Richter}}]{dembowski2004encircling}%
  \BibitemOpen
  \bibfield  {author} {\bibinfo {author} {\bibfnamefont {C.}~\bibnamefont
  {Dembowski}}, \bibinfo {author} {\bibfnamefont {B.}~\bibnamefont {Dietz}},
  \bibinfo {author} {\bibfnamefont {H.-D.}\ \bibnamefont {Gr{\"a}f}}, \bibinfo
  {author} {\bibfnamefont {H.}~\bibnamefont {Harney}}, \bibinfo {author}
  {\bibfnamefont {A.}~\bibnamefont {Heine}}, \bibinfo {author} {\bibfnamefont
  {W.}~\bibnamefont {Heiss}},\ and\ \bibinfo {author} {\bibfnamefont
  {A.}~\bibnamefont {Richter}},\ }\bibfield  {title} {\bibinfo {title}
  {Encircling an exceptional point},\ }\href@noop {} {\bibfield  {journal}
  {\bibinfo  {journal} {Physical Review E}\ }\textbf {\bibinfo {volume} {69}},\
  \bibinfo {pages} {056216} (\bibinfo {year} {2004})}\BibitemShut {NoStop}%
\bibitem [{\citenamefont {Rotter}(2009)}]{rotter2009non}%
  \BibitemOpen
  \bibfield  {author} {\bibinfo {author} {\bibfnamefont {I.}~\bibnamefont
  {Rotter}},\ }\bibfield  {title} {\bibinfo {title} {A non-hermitian hamilton
  operator and the physics of open quantum systems},\ }\href@noop {} {\bibfield
   {journal} {\bibinfo  {journal} {Journal of Physics A: Mathematical and
  Theoretical}\ }\textbf {\bibinfo {volume} {42}},\ \bibinfo {pages} {153001}
  (\bibinfo {year} {2009})}\BibitemShut {NoStop}%
\bibitem [{\citenamefont {Jin}\ and\ \citenamefont
  {Song}(2009)}]{jin2009solutions}%
  \BibitemOpen
  \bibfield  {author} {\bibinfo {author} {\bibfnamefont {L.}~\bibnamefont
  {Jin}}\ and\ \bibinfo {author} {\bibfnamefont {Z.}~\bibnamefont {Song}},\
  }\bibfield  {title} {\bibinfo {title} {Solutions of p t-symmetric
  tight-binding chain and its equivalent hermitian counterpart},\ }\href@noop
  {} {\bibfield  {journal} {\bibinfo  {journal} {Physical Review A}\ }\textbf
  {\bibinfo {volume} {80}},\ \bibinfo {pages} {052107} (\bibinfo {year}
  {2009})}\BibitemShut {NoStop}%
\bibitem [{\citenamefont {Longhi}(2010)}]{longhi2010pt}%
  \BibitemOpen
  \bibfield  {author} {\bibinfo {author} {\bibfnamefont {S.}~\bibnamefont
  {Longhi}},\ }\bibfield  {title} {\bibinfo {title} {Pt-symmetric laser
  absorber},\ }\href@noop {} {\bibfield  {journal} {\bibinfo  {journal}
  {Physical Review A}\ }\textbf {\bibinfo {volume} {82}},\ \bibinfo {pages}
  {031801} (\bibinfo {year} {2010})}\BibitemShut {NoStop}%
\bibitem [{\citenamefont {Heiss}\ and\ \citenamefont
  {Harney}(2001)}]{heiss2001chirality}%
  \BibitemOpen
  \bibfield  {author} {\bibinfo {author} {\bibfnamefont {W.}~\bibnamefont
  {Heiss}}\ and\ \bibinfo {author} {\bibfnamefont {H.}~\bibnamefont {Harney}},\
  }\bibfield  {title} {\bibinfo {title} {The chirality of exceptional points},\
  }\href@noop {} {\bibfield  {journal} {\bibinfo  {journal} {The European
  Physical Journal D-Atomic, Molecular, Optical and Plasma Physics}\ }\textbf
  {\bibinfo {volume} {17}},\ \bibinfo {pages} {149} (\bibinfo {year}
  {2001})}\BibitemShut {NoStop}%
\bibitem [{\citenamefont {Heiss}(2012)}]{heiss2012physics}%
  \BibitemOpen
  \bibfield  {author} {\bibinfo {author} {\bibfnamefont {W.}~\bibnamefont
  {Heiss}},\ }\bibfield  {title} {\bibinfo {title} {The physics of exceptional
  points},\ }\href@noop {} {\bibfield  {journal} {\bibinfo  {journal} {Journal
  of Physics A: Mathematical and Theoretical}\ }\textbf {\bibinfo {volume}
  {45}},\ \bibinfo {pages} {444016} (\bibinfo {year} {2012})}\BibitemShut
  {NoStop}%
\bibitem [{\citenamefont {Hassan}\ \emph {et~al.}(2017)\citenamefont {Hassan},
  \citenamefont {Zhen}, \citenamefont {Solja{\v{c}}i{\'c}}, \citenamefont
  {Khajavikhan},\ and\ \citenamefont
  {Christodoulides}}]{hassan2017dynamically}%
  \BibitemOpen
  \bibfield  {author} {\bibinfo {author} {\bibfnamefont {A.~U.}\ \bibnamefont
  {Hassan}}, \bibinfo {author} {\bibfnamefont {B.}~\bibnamefont {Zhen}},
  \bibinfo {author} {\bibfnamefont {M.}~\bibnamefont {Solja{\v{c}}i{\'c}}},
  \bibinfo {author} {\bibfnamefont {M.}~\bibnamefont {Khajavikhan}},\ and\
  \bibinfo {author} {\bibfnamefont {D.~N.}\ \bibnamefont {Christodoulides}},\
  }\bibfield  {title} {\bibinfo {title} {Dynamically encircling exceptional
  points: exact evolution and polarization state conversion},\ }\href@noop {}
  {\bibfield  {journal} {\bibinfo  {journal} {Physical review letters}\
  }\textbf {\bibinfo {volume} {118}},\ \bibinfo {pages} {093002} (\bibinfo
  {year} {2017})}\BibitemShut {NoStop}%
\bibitem [{\citenamefont {Xu}\ \emph {et~al.}(2016)\citenamefont {Xu},
  \citenamefont {Mason}, \citenamefont {Jiang},\ and\ \citenamefont
  {Harris}}]{xu2016topological}%
  \BibitemOpen
  \bibfield  {author} {\bibinfo {author} {\bibfnamefont {H.}~\bibnamefont
  {Xu}}, \bibinfo {author} {\bibfnamefont {D.}~\bibnamefont {Mason}}, \bibinfo
  {author} {\bibfnamefont {L.}~\bibnamefont {Jiang}},\ and\ \bibinfo {author}
  {\bibfnamefont {J.}~\bibnamefont {Harris}},\ }\bibfield  {title} {\bibinfo
  {title} {Topological energy transfer in an optomechanical system with
  exceptional points},\ }\href@noop {} {\bibfield  {journal} {\bibinfo
  {journal} {Nature}\ }\textbf {\bibinfo {volume} {537}},\ \bibinfo {pages}
  {80} (\bibinfo {year} {2016})}\BibitemShut {NoStop}%
\bibitem [{\citenamefont {Lin}\ \emph {et~al.}(2017)\citenamefont {Lin},
  \citenamefont {Hu}, \citenamefont {Chen}, \citenamefont {Lee},\ and\
  \citenamefont {Zhang}}]{lin2017line}%
  \BibitemOpen
  \bibfield  {author} {\bibinfo {author} {\bibfnamefont {J.~Y.}\ \bibnamefont
  {Lin}}, \bibinfo {author} {\bibfnamefont {N.~C.}\ \bibnamefont {Hu}},
  \bibinfo {author} {\bibfnamefont {Y.~J.}\ \bibnamefont {Chen}}, \bibinfo
  {author} {\bibfnamefont {C.~H.}\ \bibnamefont {Lee}},\ and\ \bibinfo {author}
  {\bibfnamefont {X.}~\bibnamefont {Zhang}},\ }\bibfield  {title} {\bibinfo
  {title} {Line nodes, dirac points, and lifshitz transition in two-dimensional
  nonsymmorphic photonic crystals},\ }\href@noop {} {\bibfield  {journal}
  {\bibinfo  {journal} {Physical Review B}\ }\textbf {\bibinfo {volume} {96}},\
  \bibinfo {pages} {075438} (\bibinfo {year} {2017})}\BibitemShut {NoStop}%
\bibitem [{\citenamefont {Hu}\ \emph {et~al.}(2017)\citenamefont {Hu},
  \citenamefont {Wang}, \citenamefont {Shum},\ and\ \citenamefont
  {Chong}}]{hu2017exceptional}%
  \BibitemOpen
  \bibfield  {author} {\bibinfo {author} {\bibfnamefont {W.}~\bibnamefont
  {Hu}}, \bibinfo {author} {\bibfnamefont {H.}~\bibnamefont {Wang}}, \bibinfo
  {author} {\bibfnamefont {P.~P.}\ \bibnamefont {Shum}},\ and\ \bibinfo
  {author} {\bibfnamefont {Y.~D.}\ \bibnamefont {Chong}},\ }\bibfield  {title}
  {\bibinfo {title} {Exceptional points in a non-hermitian topological pump},\
  }\href@noop {} {\bibfield  {journal} {\bibinfo  {journal} {Physical Review
  B}\ }\textbf {\bibinfo {volume} {95}},\ \bibinfo {pages} {184306} (\bibinfo
  {year} {2017})}\BibitemShut {NoStop}%
\bibitem [{\citenamefont {Shen}\ \emph {et~al.}(2018)\citenamefont {Shen},
  \citenamefont {Zhen},\ and\ \citenamefont {Fu}}]{shen2018topological}%
  \BibitemOpen
  \bibfield  {author} {\bibinfo {author} {\bibfnamefont {H.}~\bibnamefont
  {Shen}}, \bibinfo {author} {\bibfnamefont {B.}~\bibnamefont {Zhen}},\ and\
  \bibinfo {author} {\bibfnamefont {L.}~\bibnamefont {Fu}},\ }\bibfield
  {title} {\bibinfo {title} {Topological band theory for non-hermitian
  hamiltonians},\ }\href@noop {} {\bibfield  {journal} {\bibinfo  {journal}
  {Phys. Rev. Lett.}\ }\textbf {\bibinfo {volume} {120}},\ \bibinfo {pages}
  {146402} (\bibinfo {year} {2018})}\BibitemShut {NoStop}%
\bibitem [{\citenamefont {Wang}\ \emph {et~al.}(2019)\citenamefont {Wang},
  \citenamefont {Hou}, \citenamefont {Lu}, \citenamefont {Chen}, \citenamefont
  {Zhang},\ and\ \citenamefont {Chan}}]{wang2019arbitrary}%
  \BibitemOpen
  \bibfield  {author} {\bibinfo {author} {\bibfnamefont {S.}~\bibnamefont
  {Wang}}, \bibinfo {author} {\bibfnamefont {B.}~\bibnamefont {Hou}}, \bibinfo
  {author} {\bibfnamefont {W.}~\bibnamefont {Lu}}, \bibinfo {author}
  {\bibfnamefont {Y.}~\bibnamefont {Chen}}, \bibinfo {author} {\bibfnamefont
  {Z.}~\bibnamefont {Zhang}},\ and\ \bibinfo {author} {\bibfnamefont
  {C.}~\bibnamefont {Chan}},\ }\bibfield  {title} {\bibinfo {title} {Arbitrary
  order exceptional point induced by photonic spin--orbit interaction in
  coupled resonators},\ }\href@noop {} {\bibfield  {journal} {\bibinfo
  {journal} {Nature communications}\ }\textbf {\bibinfo {volume} {10}},\
  \bibinfo {pages} {1} (\bibinfo {year} {2019})}\BibitemShut {NoStop}%
\bibitem [{\citenamefont {Miri}\ and\ \citenamefont
  {Al{\`u}}(2019)}]{miri2019exceptional}%
  \BibitemOpen
  \bibfield  {author} {\bibinfo {author} {\bibfnamefont {M.-A.}\ \bibnamefont
  {Miri}}\ and\ \bibinfo {author} {\bibfnamefont {A.}~\bibnamefont {Al{\`u}}},\
  }\bibfield  {title} {\bibinfo {title} {Exceptional points in optics and
  photonics},\ }\href@noop {} {\bibfield  {journal} {\bibinfo  {journal}
  {Science}\ }\textbf {\bibinfo {volume} {363}},\ \bibinfo {pages} {eaar7709}
  (\bibinfo {year} {2019})}\BibitemShut {NoStop}%
\bibitem [{\citenamefont {Zhang}\ and\ \citenamefont
  {Gong}(2020)}]{zhang2020non}%
  \BibitemOpen
  \bibfield  {author} {\bibinfo {author} {\bibfnamefont {X.}~\bibnamefont
  {Zhang}}\ and\ \bibinfo {author} {\bibfnamefont {J.}~\bibnamefont {Gong}},\
  }\bibfield  {title} {\bibinfo {title} {Non-hermitian floquet topological
  phases: Exceptional points, coalescent edge modes, and the skin effect},\
  }\href@noop {} {\bibfield  {journal} {\bibinfo  {journal} {Physical Review
  B}\ }\textbf {\bibinfo {volume} {101}},\ \bibinfo {pages} {045415} (\bibinfo
  {year} {2020})}\BibitemShut {NoStop}%
\bibitem [{\citenamefont {Jin}\ \emph {et~al.}(2020)\citenamefont {Jin},
  \citenamefont {Wu}, \citenamefont {Wei},\ and\ \citenamefont
  {Song}}]{jin2020hybrid}%
  \BibitemOpen
  \bibfield  {author} {\bibinfo {author} {\bibfnamefont {L.}~\bibnamefont
  {Jin}}, \bibinfo {author} {\bibfnamefont {H.}~\bibnamefont {Wu}}, \bibinfo
  {author} {\bibfnamefont {B.-B.}\ \bibnamefont {Wei}},\ and\ \bibinfo {author}
  {\bibfnamefont {Z.}~\bibnamefont {Song}},\ }\bibfield  {title} {\bibinfo
  {title} {Hybrid exceptional point created from type-iii dirac point},\
  }\href@noop {} {\bibfield  {journal} {\bibinfo  {journal} {Physical Review
  B}\ }\textbf {\bibinfo {volume} {101}},\ \bibinfo {pages} {045130} (\bibinfo
  {year} {2020})}\BibitemShut {NoStop}%
\bibitem [{\citenamefont {Kawabata}\ \emph {et~al.}(2019)\citenamefont
  {Kawabata}, \citenamefont {Bessho},\ and\ \citenamefont
  {Sato}}]{kawabata2019classification}%
  \BibitemOpen
  \bibfield  {author} {\bibinfo {author} {\bibfnamefont {K.}~\bibnamefont
  {Kawabata}}, \bibinfo {author} {\bibfnamefont {T.}~\bibnamefont {Bessho}},\
  and\ \bibinfo {author} {\bibfnamefont {M.}~\bibnamefont {Sato}},\ }\bibfield
  {title} {\bibinfo {title} {Classification of exceptional points and
  non-hermitian topological semimetals},\ }\href@noop {} {\bibfield  {journal}
  {\bibinfo  {journal} {Physical review letters}\ }\textbf {\bibinfo {volume}
  {123}},\ \bibinfo {pages} {066405} (\bibinfo {year} {2019})}\BibitemShut
  {NoStop}%
\bibitem [{\citenamefont {Feng}\ \emph {et~al.}(2013)\citenamefont {Feng},
  \citenamefont {Xu}, \citenamefont {Fegadolli}, \citenamefont {Lu},
  \citenamefont {Oliveira}, \citenamefont {Almeida}, \citenamefont {Chen},\
  and\ \citenamefont {Scherer}}]{feng2013experimental}%
  \BibitemOpen
  \bibfield  {author} {\bibinfo {author} {\bibfnamefont {L.}~\bibnamefont
  {Feng}}, \bibinfo {author} {\bibfnamefont {Y.-L.}\ \bibnamefont {Xu}},
  \bibinfo {author} {\bibfnamefont {W.~S.}\ \bibnamefont {Fegadolli}}, \bibinfo
  {author} {\bibfnamefont {M.-H.}\ \bibnamefont {Lu}}, \bibinfo {author}
  {\bibfnamefont {J.~E.}\ \bibnamefont {Oliveira}}, \bibinfo {author}
  {\bibfnamefont {V.~R.}\ \bibnamefont {Almeida}}, \bibinfo {author}
  {\bibfnamefont {Y.-F.}\ \bibnamefont {Chen}},\ and\ \bibinfo {author}
  {\bibfnamefont {A.}~\bibnamefont {Scherer}},\ }\bibfield  {title} {\bibinfo
  {title} {Experimental demonstration of a unidirectional reflectionless
  parity-time metamaterial at optical frequencies},\ }\href@noop {} {\bibfield
  {journal} {\bibinfo  {journal} {Nature materials}\ }\textbf {\bibinfo
  {volume} {12}},\ \bibinfo {pages} {108} (\bibinfo {year} {2013})}\BibitemShut
  {NoStop}%
\bibitem [{\citenamefont {Liu}\ \emph {et~al.}(2018)\citenamefont {Liu},
  \citenamefont {Zhu}, \citenamefont {Chen}, \citenamefont {Liang},\ and\
  \citenamefont {Zhu}}]{liu2018unidirectional}%
  \BibitemOpen
  \bibfield  {author} {\bibinfo {author} {\bibfnamefont {T.}~\bibnamefont
  {Liu}}, \bibinfo {author} {\bibfnamefont {X.}~\bibnamefont {Zhu}}, \bibinfo
  {author} {\bibfnamefont {F.}~\bibnamefont {Chen}}, \bibinfo {author}
  {\bibfnamefont {S.}~\bibnamefont {Liang}},\ and\ \bibinfo {author}
  {\bibfnamefont {J.}~\bibnamefont {Zhu}},\ }\bibfield  {title} {\bibinfo
  {title} {Unidirectional wave vector manipulation in two-dimensional space
  with an all passive acoustic parity-time-symmetric metamaterials crystal},\
  }\href@noop {} {\bibfield  {journal} {\bibinfo  {journal} {Physical review
  letters}\ }\textbf {\bibinfo {volume} {120}},\ \bibinfo {pages} {124502}
  (\bibinfo {year} {2018})}\BibitemShut {NoStop}%
\bibitem [{\citenamefont {Almheiri}\ \emph {et~al.}(2021)\citenamefont
  {Almheiri}, \citenamefont {Hartman}, \citenamefont {Maldacena}, \citenamefont
  {Shaghoulian},\ and\ \citenamefont {Tajdini}}]{almheiri2020entropy}%
  \BibitemOpen
  \bibfield  {author} {\bibinfo {author} {\bibfnamefont {A.}~\bibnamefont
  {Almheiri}}, \bibinfo {author} {\bibfnamefont {T.}~\bibnamefont {Hartman}},
  \bibinfo {author} {\bibfnamefont {J.}~\bibnamefont {Maldacena}}, \bibinfo
  {author} {\bibfnamefont {E.}~\bibnamefont {Shaghoulian}},\ and\ \bibinfo
  {author} {\bibfnamefont {A.}~\bibnamefont {Tajdini}},\ }\bibfield  {title}
  {\bibinfo {title} {The entropy of hawking radiation},\ }\href@noop {}
  {\bibfield  {journal} {\bibinfo  {journal} {Reviews of Modern Physics}\
  }\textbf {\bibinfo {volume} {93}},\ \bibinfo {pages} {035002} (\bibinfo
  {year} {2021})}\BibitemShut {NoStop}%
\bibitem [{\citenamefont {Chen}\ \emph {et~al.}(2020)\citenamefont {Chen},
  \citenamefont {Myers}, \citenamefont {Neuenfeld}, \citenamefont {Reyes},\
  and\ \citenamefont {Sandor}}]{chen2020quantum}%
  \BibitemOpen
  \bibfield  {author} {\bibinfo {author} {\bibfnamefont {H.~Z.}\ \bibnamefont
  {Chen}}, \bibinfo {author} {\bibfnamefont {R.~C.}\ \bibnamefont {Myers}},
  \bibinfo {author} {\bibfnamefont {D.}~\bibnamefont {Neuenfeld}}, \bibinfo
  {author} {\bibfnamefont {I.~A.}\ \bibnamefont {Reyes}},\ and\ \bibinfo
  {author} {\bibfnamefont {J.}~\bibnamefont {Sandor}},\ }\bibfield  {title}
  {\bibinfo {title} {Quantum extremal islands made easy. part i. entanglement
  on the brane},\ }\href@noop {} {\bibfield  {journal} {\bibinfo  {journal}
  {Journal of High Energy Physics}\ }\textbf {\bibinfo {volume} {2020}},\
  \bibinfo {pages} {1} (\bibinfo {year} {2020})}\BibitemShut {NoStop}%
\bibitem [{\citenamefont {Qi}(2011)}]{qi2011generic}%
  \BibitemOpen
  \bibfield  {author} {\bibinfo {author} {\bibfnamefont {X.-L.}\ \bibnamefont
  {Qi}},\ }\bibfield  {title} {\bibinfo {title} {Generic wave-function
  description of fractional quantum anomalous hall states and fractional
  topological insulators},\ }\href@noop {} {\bibfield  {journal} {\bibinfo
  {journal} {Phys. Rev. Lett.}\ }\textbf {\bibinfo {volume} {107}},\ \bibinfo
  {pages} {126803} (\bibinfo {year} {2011})}\BibitemShut {NoStop}%
\bibitem [{\citenamefont {Alexandradinata}\ \emph {et~al.}(2011)\citenamefont
  {Alexandradinata}, \citenamefont {Hughes},\ and\ \citenamefont
  {Bernevig}}]{alexandradinata2011trace}%
  \BibitemOpen
  \bibfield  {author} {\bibinfo {author} {\bibfnamefont {A.}~\bibnamefont
  {Alexandradinata}}, \bibinfo {author} {\bibfnamefont {T.~L.}\ \bibnamefont
  {Hughes}},\ and\ \bibinfo {author} {\bibfnamefont {B.~A.}\ \bibnamefont
  {Bernevig}},\ }\bibfield  {title} {\bibinfo {title} {Trace index and spectral
  flow in the entanglement spectrum of topological insulators},\ }\href@noop {}
  {\bibfield  {journal} {\bibinfo  {journal} {Phys. Rev. B}\ }\textbf {\bibinfo
  {volume} {84}},\ \bibinfo {pages} {195103} (\bibinfo {year}
  {2011})}\BibitemShut {NoStop}%
\bibitem [{\citenamefont {Yu}\ \emph {et~al.}(2011)\citenamefont {Yu},
  \citenamefont {Qi}, \citenamefont {Bernevig}, \citenamefont {Fang},\ and\
  \citenamefont {Dai}}]{yu2011equivalent}%
  \BibitemOpen
  \bibfield  {author} {\bibinfo {author} {\bibfnamefont {R.}~\bibnamefont
  {Yu}}, \bibinfo {author} {\bibfnamefont {X.~L.}\ \bibnamefont {Qi}}, \bibinfo
  {author} {\bibfnamefont {A.}~\bibnamefont {Bernevig}}, \bibinfo {author}
  {\bibfnamefont {Z.}~\bibnamefont {Fang}},\ and\ \bibinfo {author}
  {\bibfnamefont {X.}~\bibnamefont {Dai}},\ }\bibfield  {title} {\bibinfo
  {title} {Equivalent expression of z 2 topological invariant for band
  insulators using the non-abelian berry connection},\ }\href@noop {}
  {\bibfield  {journal} {\bibinfo  {journal} {Physical Review B}\ }\textbf
  {\bibinfo {volume} {84}},\ \bibinfo {pages} {075119} (\bibinfo {year}
  {2011})}\BibitemShut {NoStop}%
\bibitem [{\citenamefont {Lee}\ and\ \citenamefont {Ye}(2015)}]{lee2015free}%
  \BibitemOpen
  \bibfield  {author} {\bibinfo {author} {\bibfnamefont {C.~H.}\ \bibnamefont
  {Lee}}\ and\ \bibinfo {author} {\bibfnamefont {P.}~\bibnamefont {Ye}},\
  }\bibfield  {title} {\bibinfo {title} {Free-fermion entanglement spectrum
  through wannier interpolation},\ }\href@noop {} {\bibfield  {journal}
  {\bibinfo  {journal} {Physical Review B}\ }\textbf {\bibinfo {volume} {91}},\
  \bibinfo {pages} {085119} (\bibinfo {year} {2015})}\BibitemShut {NoStop}%
\bibitem [{\citenamefont {Kim}\ \emph {et~al.}(2021)\citenamefont {Kim},
  \citenamefont {Zhang}, \citenamefont {Ferreira}, \citenamefont {Banker},
  \citenamefont {Iverson}, \citenamefont {Sipahigil}, \citenamefont {Bello},
  \citenamefont {Gonz{\'a}lez-Tudela}, \citenamefont {Mirhosseini},\ and\
  \citenamefont {Painter}}]{kim2021quantum}%
  \BibitemOpen
  \bibfield  {author} {\bibinfo {author} {\bibfnamefont {E.}~\bibnamefont
  {Kim}}, \bibinfo {author} {\bibfnamefont {X.}~\bibnamefont {Zhang}}, \bibinfo
  {author} {\bibfnamefont {V.~S.}\ \bibnamefont {Ferreira}}, \bibinfo {author}
  {\bibfnamefont {J.}~\bibnamefont {Banker}}, \bibinfo {author} {\bibfnamefont
  {J.~K.}\ \bibnamefont {Iverson}}, \bibinfo {author} {\bibfnamefont
  {A.}~\bibnamefont {Sipahigil}}, \bibinfo {author} {\bibfnamefont
  {M.}~\bibnamefont {Bello}}, \bibinfo {author} {\bibfnamefont
  {A.}~\bibnamefont {Gonz{\'a}lez-Tudela}}, \bibinfo {author} {\bibfnamefont
  {M.}~\bibnamefont {Mirhosseini}},\ and\ \bibinfo {author} {\bibfnamefont
  {O.}~\bibnamefont {Painter}},\ }\bibfield  {title} {\bibinfo {title} {Quantum
  electrodynamics in a topological waveguide},\ }\href@noop {} {\bibfield
  {journal} {\bibinfo  {journal} {Physical Review X}\ }\textbf {\bibinfo
  {volume} {11}},\ \bibinfo {pages} {011015} (\bibinfo {year}
  {2021})}\BibitemShut {NoStop}%
\bibitem [{\citenamefont {Weimann}\ \emph {et~al.}(2017)\citenamefont
  {Weimann}, \citenamefont {Kremer}, \citenamefont {Plotnik}, \citenamefont
  {Lumer}, \citenamefont {Nolte}, \citenamefont {Makris}, \citenamefont
  {Segev}, \citenamefont {Rechtsman},\ and\ \citenamefont
  {Szameit}}]{weimann2017topologically}%
  \BibitemOpen
  \bibfield  {author} {\bibinfo {author} {\bibfnamefont {S.}~\bibnamefont
  {Weimann}}, \bibinfo {author} {\bibfnamefont {M.}~\bibnamefont {Kremer}},
  \bibinfo {author} {\bibfnamefont {Y.}~\bibnamefont {Plotnik}}, \bibinfo
  {author} {\bibfnamefont {Y.}~\bibnamefont {Lumer}}, \bibinfo {author}
  {\bibfnamefont {S.}~\bibnamefont {Nolte}}, \bibinfo {author} {\bibfnamefont
  {K.~G.}\ \bibnamefont {Makris}}, \bibinfo {author} {\bibfnamefont
  {M.}~\bibnamefont {Segev}}, \bibinfo {author} {\bibfnamefont {M.~C.}\
  \bibnamefont {Rechtsman}},\ and\ \bibinfo {author} {\bibfnamefont
  {A.}~\bibnamefont {Szameit}},\ }\bibfield  {title} {\bibinfo {title}
  {Topologically protected bound states in photonic parity--time-symmetric
  crystals},\ }\href@noop {} {\bibfield  {journal} {\bibinfo  {journal} {Nature
  materials}\ }\textbf {\bibinfo {volume} {16}},\ \bibinfo {pages} {433}
  (\bibinfo {year} {2017})}\BibitemShut {NoStop}%
\bibitem [{\citenamefont {Budich}\ and\ \citenamefont
  {Bergholtz}(2020)}]{budich2020non}%
  \BibitemOpen
  \bibfield  {author} {\bibinfo {author} {\bibfnamefont {J.~C.}\ \bibnamefont
  {Budich}}\ and\ \bibinfo {author} {\bibfnamefont {E.~J.}\ \bibnamefont
  {Bergholtz}},\ }\bibfield  {title} {\bibinfo {title} {Non-hermitian
  topological sensors},\ }\href@noop {} {\bibfield  {journal} {\bibinfo
  {journal} {Physical Review Letters}\ }\textbf {\bibinfo {volume} {125}},\
  \bibinfo {pages} {180403} (\bibinfo {year} {2020})}\BibitemShut {NoStop}%
\bibitem [{\citenamefont {Hsu}\ \emph {et~al.}(2016)\citenamefont {Hsu},
  \citenamefont {Zhen}, \citenamefont {Stone}, \citenamefont {Joannopoulos},\
  and\ \citenamefont {Solja{\v{c}}i{\'c}}}]{hsu2016bound}%
  \BibitemOpen
  \bibfield  {author} {\bibinfo {author} {\bibfnamefont {C.~W.}\ \bibnamefont
  {Hsu}}, \bibinfo {author} {\bibfnamefont {B.}~\bibnamefont {Zhen}}, \bibinfo
  {author} {\bibfnamefont {A.~D.}\ \bibnamefont {Stone}}, \bibinfo {author}
  {\bibfnamefont {J.~D.}\ \bibnamefont {Joannopoulos}},\ and\ \bibinfo {author}
  {\bibfnamefont {M.}~\bibnamefont {Solja{\v{c}}i{\'c}}},\ }\bibfield  {title}
  {\bibinfo {title} {Bound states in the continuum},\ }\href@noop {} {\bibfield
   {journal} {\bibinfo  {journal} {Nature Reviews Materials}\ }\textbf
  {\bibinfo {volume} {1}},\ \bibinfo {pages} {1} (\bibinfo {year}
  {2016})}\BibitemShut {NoStop}%
\bibitem [{\citenamefont {Zhen}\ \emph {et~al.}(2014)\citenamefont {Zhen},
  \citenamefont {Hsu}, \citenamefont {Lu}, \citenamefont {Stone},\ and\
  \citenamefont {Solja{\v{c}}i{\'c}}}]{zhen2014topological}%
  \BibitemOpen
  \bibfield  {author} {\bibinfo {author} {\bibfnamefont {B.}~\bibnamefont
  {Zhen}}, \bibinfo {author} {\bibfnamefont {C.~W.}\ \bibnamefont {Hsu}},
  \bibinfo {author} {\bibfnamefont {L.}~\bibnamefont {Lu}}, \bibinfo {author}
  {\bibfnamefont {A.~D.}\ \bibnamefont {Stone}},\ and\ \bibinfo {author}
  {\bibfnamefont {M.}~\bibnamefont {Solja{\v{c}}i{\'c}}},\ }\bibfield  {title}
  {\bibinfo {title} {Topological nature of optical bound states in the
  continuum},\ }\href@noop {} {\bibfield  {journal} {\bibinfo  {journal}
  {Physical review letters}\ }\textbf {\bibinfo {volume} {113}},\ \bibinfo
  {pages} {257401} (\bibinfo {year} {2014})}\BibitemShut {NoStop}%
\bibitem [{\citenamefont {Morvan}\ \emph {et~al.}(2022)\citenamefont {Morvan},
  \citenamefont {Andersen}, \citenamefont {Mi}, \citenamefont {Neill},
  \citenamefont {Petukhov}, \citenamefont {Kechedzhi}, \citenamefont {Abanin},
  \citenamefont {Michailidis}, \citenamefont {Acharya}, \citenamefont {Arute}
  \emph {et~al.}}]{morvan2022formation}%
  \BibitemOpen
  \bibfield  {author} {\bibinfo {author} {\bibfnamefont {A.}~\bibnamefont
  {Morvan}}, \bibinfo {author} {\bibfnamefont {T.}~\bibnamefont {Andersen}},
  \bibinfo {author} {\bibfnamefont {X.}~\bibnamefont {Mi}}, \bibinfo {author}
  {\bibfnamefont {C.}~\bibnamefont {Neill}}, \bibinfo {author} {\bibfnamefont
  {A.}~\bibnamefont {Petukhov}}, \bibinfo {author} {\bibfnamefont
  {K.}~\bibnamefont {Kechedzhi}}, \bibinfo {author} {\bibfnamefont
  {D.}~\bibnamefont {Abanin}}, \bibinfo {author} {\bibfnamefont
  {A.}~\bibnamefont {Michailidis}}, \bibinfo {author} {\bibfnamefont
  {R.}~\bibnamefont {Acharya}}, \bibinfo {author} {\bibfnamefont
  {F.}~\bibnamefont {Arute}}, \emph {et~al.},\ }\bibfield  {title} {\bibinfo
  {title} {Formation of robust bound states of interacting microwave photons},\
  }\href@noop {} {\bibfield  {journal} {\bibinfo  {journal} {Nature}\ }\textbf
  {\bibinfo {volume} {612}},\ \bibinfo {pages} {240} (\bibinfo {year}
  {2022})}\BibitemShut {NoStop}%
\bibitem [{\citenamefont {Haldane}(1988)}]{haldane1988model}%
  \BibitemOpen
  \bibfield  {author} {\bibinfo {author} {\bibfnamefont {F.~D.~M.}\
  \bibnamefont {Haldane}},\ }\bibfield  {title} {\bibinfo {title} {Model for a
  quantum hall effect without landau levels: Condensed-matter realization of
  the" parity anomaly"},\ }\href@noop {} {\bibfield  {journal} {\bibinfo
  {journal} {Phys. Rev. Lett.}\ }\textbf {\bibinfo {volume} {61}},\ \bibinfo
  {pages} {2015} (\bibinfo {year} {1988})}\BibitemShut {NoStop}%
\bibitem [{\citenamefont {Kane}\ and\ \citenamefont {Mele}(2005)}]{kane2005z}%
  \BibitemOpen
  \bibfield  {author} {\bibinfo {author} {\bibfnamefont {C.~L.}\ \bibnamefont
  {Kane}}\ and\ \bibinfo {author} {\bibfnamefont {E.~J.}\ \bibnamefont
  {Mele}},\ }\bibfield  {title} {\bibinfo {title} {Z 2 topological order and
  the quantum spin hall effect},\ }\href@noop {} {\bibfield  {journal}
  {\bibinfo  {journal} {Phys. Rev. Lett.}\ }\textbf {\bibinfo {volume} {95}},\
  \bibinfo {pages} {146802} (\bibinfo {year} {2005})}\BibitemShut {NoStop}%
\bibitem [{\citenamefont {Schnyder}\ \emph {et~al.}(2009)\citenamefont
  {Schnyder}, \citenamefont {Ryu}, \citenamefont {Furusaki},\ and\
  \citenamefont {Ludwig}}]{schnyder2009classification}%
  \BibitemOpen
  \bibfield  {author} {\bibinfo {author} {\bibfnamefont {A.~P.}\ \bibnamefont
  {Schnyder}}, \bibinfo {author} {\bibfnamefont {S.}~\bibnamefont {Ryu}},
  \bibinfo {author} {\bibfnamefont {A.}~\bibnamefont {Furusaki}},\ and\
  \bibinfo {author} {\bibfnamefont {A.~W.}\ \bibnamefont {Ludwig}},\ }\bibfield
   {title} {\bibinfo {title} {Classification of topological insulators and
  superconductors},\ }in\ \href@noop {} {\emph {\bibinfo {booktitle} {AIP
  Conference Proceedings}}},\ Vol.\ \bibinfo {volume} {1134}\ (\bibinfo
  {organization} {American Institute of Physics},\ \bibinfo {year} {2009})\
  pp.\ \bibinfo {pages} {10--21}\BibitemShut {NoStop}%
\bibitem [{\citenamefont {Chang}\ \emph {et~al.}(2023)\citenamefont {Chang},
  \citenamefont {Liu},\ and\ \citenamefont {MacDonald}}]{chang2023colloquium}%
  \BibitemOpen
  \bibfield  {author} {\bibinfo {author} {\bibfnamefont {C.-Z.}\ \bibnamefont
  {Chang}}, \bibinfo {author} {\bibfnamefont {C.-X.}\ \bibnamefont {Liu}},\
  and\ \bibinfo {author} {\bibfnamefont {A.~H.}\ \bibnamefont {MacDonald}},\
  }\bibfield  {title} {\bibinfo {title} {Colloquium: Quantum anomalous hall
  effect},\ }\href@noop {} {\bibfield  {journal} {\bibinfo  {journal} {Reviews
  of Modern Physics}\ }\textbf {\bibinfo {volume} {95}},\ \bibinfo {pages}
  {011002} (\bibinfo {year} {2023})}\BibitemShut {NoStop}%
\bibitem [{\citenamefont {Yao}\ and\ \citenamefont {Wang}(2018)}]{yao2018edge}%
  \BibitemOpen
  \bibfield  {author} {\bibinfo {author} {\bibfnamefont {S.}~\bibnamefont
  {Yao}}\ and\ \bibinfo {author} {\bibfnamefont {Z.}~\bibnamefont {Wang}},\
  }\bibfield  {title} {\bibinfo {title} {Edge states and topological invariants
  of non-hermitian systems},\ }\href
  {https://doi.org/10.1103/PhysRevLett.121.086803} {\bibfield  {journal}
  {\bibinfo  {journal} {Phys. Rev. Lett.}\ }\textbf {\bibinfo {volume} {121}},\
  \bibinfo {pages} {086803} (\bibinfo {year} {2018})}\BibitemShut {NoStop}%
\bibitem [{\citenamefont {Xiong}(2018)}]{xiong2018does}%
  \BibitemOpen
  \bibfield  {author} {\bibinfo {author} {\bibfnamefont {Y.}~\bibnamefont
  {Xiong}},\ }\bibfield  {title} {\bibinfo {title} {Why does bulk boundary
  correspondence fail in some non-hermitian topological models},\ }\href@noop
  {} {\bibfield  {journal} {\bibinfo  {journal} {Journal of Physics
  Communications}\ }\textbf {\bibinfo {volume} {2}},\ \bibinfo {pages} {035043}
  (\bibinfo {year} {2018})}\BibitemShut {NoStop}%
\bibitem [{\citenamefont {Lee}\ and\ \citenamefont
  {Thomale}(2019)}]{lee2019anatomy}%
  \BibitemOpen
  \bibfield  {author} {\bibinfo {author} {\bibfnamefont {C.~H.}\ \bibnamefont
  {Lee}}\ and\ \bibinfo {author} {\bibfnamefont {R.}~\bibnamefont {Thomale}},\
  }\bibfield  {title} {\bibinfo {title} {Anatomy of skin modes and topology in
  non-hermitian systems},\ }\href {https://doi.org/10.1103/PhysRevB.99.201103}
  {\bibfield  {journal} {\bibinfo  {journal} {Phys. Rev. B}\ }\textbf {\bibinfo
  {volume} {99}},\ \bibinfo {pages} {201103} (\bibinfo {year}
  {2019})}\BibitemShut {NoStop}%
\bibitem [{\citenamefont {Yokomizo}\ and\ \citenamefont
  {Murakami}(2019)}]{yokomizo2019non}%
  \BibitemOpen
  \bibfield  {author} {\bibinfo {author} {\bibfnamefont {K.}~\bibnamefont
  {Yokomizo}}\ and\ \bibinfo {author} {\bibfnamefont {S.}~\bibnamefont
  {Murakami}},\ }\bibfield  {title} {\bibinfo {title} {Non-bloch band theory of
  non-hermitian systems},\ }\href@noop {} {\bibfield  {journal} {\bibinfo
  {journal} {Physical review letters}\ }\textbf {\bibinfo {volume} {123}},\
  \bibinfo {pages} {066404} (\bibinfo {year} {2019})}\BibitemShut {NoStop}%
\bibitem [{\citenamefont {Kunst}\ \emph {et~al.}(2018)\citenamefont {Kunst},
  \citenamefont {Edvardsson}, \citenamefont {Budich},\ and\ \citenamefont
  {Bergholtz}}]{kunst2018biorthogonal}%
  \BibitemOpen
  \bibfield  {author} {\bibinfo {author} {\bibfnamefont {F.~K.}\ \bibnamefont
  {Kunst}}, \bibinfo {author} {\bibfnamefont {E.}~\bibnamefont {Edvardsson}},
  \bibinfo {author} {\bibfnamefont {J.~C.}\ \bibnamefont {Budich}},\ and\
  \bibinfo {author} {\bibfnamefont {E.~J.}\ \bibnamefont {Bergholtz}},\
  }\bibfield  {title} {\bibinfo {title} {Biorthogonal bulk-boundary
  correspondence in non-hermitian systems},\ }\href
  {https://doi.org/10.1103/PhysRevLett.121.026808} {\bibfield  {journal}
  {\bibinfo  {journal} {Phys. Rev. Lett.}\ }\textbf {\bibinfo {volume} {121}},\
  \bibinfo {pages} {026808} (\bibinfo {year} {2018})}\BibitemShut {NoStop}%
\bibitem [{\citenamefont {Lee}\ \emph {et~al.}(2020{\natexlab{a}})\citenamefont
  {Lee}, \citenamefont {Li}, \citenamefont {Thomale},\ and\ \citenamefont
  {Gong}}]{lee2020unraveling}%
  \BibitemOpen
  \bibfield  {author} {\bibinfo {author} {\bibfnamefont {C.~H.}\ \bibnamefont
  {Lee}}, \bibinfo {author} {\bibfnamefont {L.}~\bibnamefont {Li}}, \bibinfo
  {author} {\bibfnamefont {R.}~\bibnamefont {Thomale}},\ and\ \bibinfo {author}
  {\bibfnamefont {J.}~\bibnamefont {Gong}},\ }\bibfield  {title} {\bibinfo
  {title} {Unraveling non-hermitian pumping: Emergent spectral singularities
  and anomalous responses},\ }\href
  {https://doi.org/10.1103/PhysRevB.102.085151} {\bibfield  {journal} {\bibinfo
   {journal} {Phys. Rev. B}\ }\textbf {\bibinfo {volume} {102}},\ \bibinfo
  {pages} {085151} (\bibinfo {year} {2020}{\natexlab{a}})}\BibitemShut
  {NoStop}%
\bibitem [{\citenamefont {Peschel}\ and\ \citenamefont
  {Eisler}(2009)}]{peschel2009reduced}%
  \BibitemOpen
  \bibfield  {author} {\bibinfo {author} {\bibfnamefont {I.}~\bibnamefont
  {Peschel}}\ and\ \bibinfo {author} {\bibfnamefont {V.}~\bibnamefont
  {Eisler}},\ }\bibfield  {title} {\bibinfo {title} {Reduced density matrices
  and entanglement entropy in free lattice models},\ }\href@noop {} {\bibfield
  {journal} {\bibinfo  {journal} {Journal of physics a: mathematical and
  theoretical}\ }\textbf {\bibinfo {volume} {42}},\ \bibinfo {pages} {504003}
  (\bibinfo {year} {2009})}\BibitemShut {NoStop}%
\bibitem [{\citenamefont {Pollmann}\ \emph {et~al.}(2010)\citenamefont
  {Pollmann}, \citenamefont {Turner}, \citenamefont {Berg},\ and\ \citenamefont
  {Oshikawa}}]{pollmann2010entanglement}%
  \BibitemOpen
  \bibfield  {author} {\bibinfo {author} {\bibfnamefont {F.}~\bibnamefont
  {Pollmann}}, \bibinfo {author} {\bibfnamefont {A.~M.}\ \bibnamefont
  {Turner}}, \bibinfo {author} {\bibfnamefont {E.}~\bibnamefont {Berg}},\ and\
  \bibinfo {author} {\bibfnamefont {M.}~\bibnamefont {Oshikawa}},\ }\bibfield
  {title} {\bibinfo {title} {Entanglement spectrum of a topological phase in
  one dimension},\ }\href@noop {} {\bibfield  {journal} {\bibinfo  {journal}
  {Physical review b}\ }\textbf {\bibinfo {volume} {81}},\ \bibinfo {pages}
  {064439} (\bibinfo {year} {2010})}\BibitemShut {NoStop}%
\bibitem [{\citenamefont {Brody}(2013)}]{brody2013biorthogonal}%
  \BibitemOpen
  \bibfield  {author} {\bibinfo {author} {\bibfnamefont {D.~C.}\ \bibnamefont
  {Brody}},\ }\bibfield  {title} {\bibinfo {title} {Biorthogonal quantum
  mechanics},\ }\href@noop {} {\bibfield  {journal} {\bibinfo  {journal}
  {Journal of Physics A: Mathematical and Theoretical}\ }\textbf {\bibinfo
  {volume} {47}},\ \bibinfo {pages} {035305} (\bibinfo {year}
  {2013})}\BibitemShut {NoStop}%
\bibitem [{\citenamefont {Calabrese}\ \emph {et~al.}(2012)\citenamefont
  {Calabrese}, \citenamefont {Mintchev},\ and\ \citenamefont
  {Vicari}}]{calabrese2012entanglement}%
  \BibitemOpen
  \bibfield  {author} {\bibinfo {author} {\bibfnamefont {P.}~\bibnamefont
  {Calabrese}}, \bibinfo {author} {\bibfnamefont {M.}~\bibnamefont
  {Mintchev}},\ and\ \bibinfo {author} {\bibfnamefont {E.}~\bibnamefont
  {Vicari}},\ }\bibfield  {title} {\bibinfo {title} {Entanglement entropies in
  free-fermion gases for arbitrary dimension},\ }\href@noop {} {\bibfield
  {journal} {\bibinfo  {journal} {Europhysics Letters}\ }\textbf {\bibinfo
  {volume} {97}},\ \bibinfo {pages} {20009} (\bibinfo {year}
  {2012})}\BibitemShut {NoStop}%
\bibitem [{\citenamefont {Gioev}\ and\ \citenamefont
  {Klich}(2006)}]{gioev2006entanglement}%
  \BibitemOpen
  \bibfield  {author} {\bibinfo {author} {\bibfnamefont {D.}~\bibnamefont
  {Gioev}}\ and\ \bibinfo {author} {\bibfnamefont {I.}~\bibnamefont {Klich}},\
  }\bibfield  {title} {\bibinfo {title} {Entanglement entropy of fermions in
  any dimension and the widom conjecture},\ }\href@noop {} {\bibfield
  {journal} {\bibinfo  {journal} {Physical review letters}\ }\textbf {\bibinfo
  {volume} {96}},\ \bibinfo {pages} {100503} (\bibinfo {year}
  {2006})}\BibitemShut {NoStop}%
\bibitem [{\citenamefont {Guo}\ \emph {et~al.}(2021)\citenamefont {Guo},
  \citenamefont {Yu}, \citenamefont {Huang}, \citenamefont {Yang},
  \citenamefont {Chi}, \citenamefont {Liao},\ and\ \citenamefont
  {Xiang}}]{guo2021entanglement}%
  \BibitemOpen
  \bibfield  {author} {\bibinfo {author} {\bibfnamefont {Y.-B.}\ \bibnamefont
  {Guo}}, \bibinfo {author} {\bibfnamefont {Y.-C.}\ \bibnamefont {Yu}},
  \bibinfo {author} {\bibfnamefont {R.-Z.}\ \bibnamefont {Huang}}, \bibinfo
  {author} {\bibfnamefont {L.-P.}\ \bibnamefont {Yang}}, \bibinfo {author}
  {\bibfnamefont {R.-Z.}\ \bibnamefont {Chi}}, \bibinfo {author} {\bibfnamefont
  {H.-J.}\ \bibnamefont {Liao}},\ and\ \bibinfo {author} {\bibfnamefont
  {T.}~\bibnamefont {Xiang}},\ }\bibfield  {title} {\bibinfo {title}
  {Entanglement entropy of non-hermitian free fermions},\ }\href@noop {}
  {\bibfield  {journal} {\bibinfo  {journal} {Journal of Physics: Condensed
  Matter}\ }\textbf {\bibinfo {volume} {33}},\ \bibinfo {pages} {475502}
  (\bibinfo {year} {2021})}\BibitemShut {NoStop}%
\bibitem [{\citenamefont {Niu}\ \emph {et~al.}(1985)\citenamefont {Niu},
  \citenamefont {Thouless},\ and\ \citenamefont {Wu}}]{niu1985quantized}%
  \BibitemOpen
  \bibfield  {author} {\bibinfo {author} {\bibfnamefont {Q.}~\bibnamefont
  {Niu}}, \bibinfo {author} {\bibfnamefont {D.~J.}\ \bibnamefont {Thouless}},\
  and\ \bibinfo {author} {\bibfnamefont {Y.-S.}\ \bibnamefont {Wu}},\
  }\bibfield  {title} {\bibinfo {title} {Quantized hall conductance as a
  topological invariant},\ }\href@noop {} {\bibfield  {journal} {\bibinfo
  {journal} {Physical Review B}\ }\textbf {\bibinfo {volume} {31}},\ \bibinfo
  {pages} {3372} (\bibinfo {year} {1985})}\BibitemShut {NoStop}%
\bibitem [{\citenamefont {Lee}\ \emph {et~al.}(2018)\citenamefont {Lee},
  \citenamefont {Imhof}, \citenamefont {Berger}, \citenamefont {Bayer},
  \citenamefont {Brehm}, \citenamefont {Molenkamp}, \citenamefont {Kiessling},\
  and\ \citenamefont {Thomale}}]{lee2018topolectrical}%
  \BibitemOpen
  \bibfield  {author} {\bibinfo {author} {\bibfnamefont {C.~H.}\ \bibnamefont
  {Lee}}, \bibinfo {author} {\bibfnamefont {S.}~\bibnamefont {Imhof}}, \bibinfo
  {author} {\bibfnamefont {C.}~\bibnamefont {Berger}}, \bibinfo {author}
  {\bibfnamefont {F.}~\bibnamefont {Bayer}}, \bibinfo {author} {\bibfnamefont
  {J.}~\bibnamefont {Brehm}}, \bibinfo {author} {\bibfnamefont {L.~W.}\
  \bibnamefont {Molenkamp}}, \bibinfo {author} {\bibfnamefont {T.}~\bibnamefont
  {Kiessling}},\ and\ \bibinfo {author} {\bibfnamefont {R.}~\bibnamefont
  {Thomale}},\ }\bibfield  {title} {\bibinfo {title} {Topolectrical circuits},\
  }\href@noop {} {\bibfield  {journal} {\bibinfo  {journal} {Communications
  Physics}\ }\textbf {\bibinfo {volume} {1}},\ \bibinfo {pages} {1} (\bibinfo
  {year} {2018})}\BibitemShut {NoStop}%
\bibitem [{\citenamefont {Lee}\ \emph {et~al.}(2020{\natexlab{b}})\citenamefont
  {Lee}, \citenamefont {Sutrisno}, \citenamefont {Hofmann}, \citenamefont
  {Helbig}, \citenamefont {Liu}, \citenamefont {Ang}, \citenamefont {Ang},
  \citenamefont {Zhang}, \citenamefont {Greiter},\ and\ \citenamefont
  {Thomale}}]{lee2020imaging}%
  \BibitemOpen
  \bibfield  {author} {\bibinfo {author} {\bibfnamefont {C.~H.}\ \bibnamefont
  {Lee}}, \bibinfo {author} {\bibfnamefont {A.}~\bibnamefont {Sutrisno}},
  \bibinfo {author} {\bibfnamefont {T.}~\bibnamefont {Hofmann}}, \bibinfo
  {author} {\bibfnamefont {T.}~\bibnamefont {Helbig}}, \bibinfo {author}
  {\bibfnamefont {Y.}~\bibnamefont {Liu}}, \bibinfo {author} {\bibfnamefont
  {Y.~S.}\ \bibnamefont {Ang}}, \bibinfo {author} {\bibfnamefont {L.~K.}\
  \bibnamefont {Ang}}, \bibinfo {author} {\bibfnamefont {X.}~\bibnamefont
  {Zhang}}, \bibinfo {author} {\bibfnamefont {M.}~\bibnamefont {Greiter}},\
  and\ \bibinfo {author} {\bibfnamefont {R.}~\bibnamefont {Thomale}},\
  }\bibfield  {title} {\bibinfo {title} {Imaging nodal knots in momentum space
  through topolectrical circuits},\ }\href@noop {} {\bibfield  {journal}
  {\bibinfo  {journal} {Nature communications}\ }\textbf {\bibinfo {volume}
  {11}},\ \bibinfo {pages} {4385} (\bibinfo {year}
  {2020}{\natexlab{b}})}\BibitemShut {NoStop}%
\bibitem [{\citenamefont {Helbig}\ \emph {et~al.}(2020)\citenamefont {Helbig},
  \citenamefont {Hofmann}, \citenamefont {Imhof}, \citenamefont {Abdelghany},
  \citenamefont {Kiessling}, \citenamefont {Molenkamp}, \citenamefont {Lee},
  \citenamefont {Szameit}, \citenamefont {Greiter},\ and\ \citenamefont
  {Thomale}}]{helbig2020generalized}%
  \BibitemOpen
  \bibfield  {author} {\bibinfo {author} {\bibfnamefont {T.}~\bibnamefont
  {Helbig}}, \bibinfo {author} {\bibfnamefont {T.}~\bibnamefont {Hofmann}},
  \bibinfo {author} {\bibfnamefont {S.}~\bibnamefont {Imhof}}, \bibinfo
  {author} {\bibfnamefont {M.}~\bibnamefont {Abdelghany}}, \bibinfo {author}
  {\bibfnamefont {T.}~\bibnamefont {Kiessling}}, \bibinfo {author}
  {\bibfnamefont {L.}~\bibnamefont {Molenkamp}}, \bibinfo {author}
  {\bibfnamefont {C.}~\bibnamefont {Lee}}, \bibinfo {author} {\bibfnamefont
  {A.}~\bibnamefont {Szameit}}, \bibinfo {author} {\bibfnamefont
  {M.}~\bibnamefont {Greiter}},\ and\ \bibinfo {author} {\bibfnamefont
  {R.}~\bibnamefont {Thomale}},\ }\bibfield  {title} {\bibinfo {title}
  {Generalized bulk-boundary correspondence in non-hermitian topolectrical
  circuits},\ }\href@noop {} {\bibfield  {journal} {\bibinfo  {journal} {Nature
  Physics}\ }\textbf {\bibinfo {volume} {16}},\ \bibinfo {pages} {747–750}
  (\bibinfo {year} {2020})}\BibitemShut {NoStop}%
\end{thebibliography}%

\clearpage

\onecolumngrid
\begin{center}
\textbf{\large Supplementary materials}\end{center}
\setcounter{equation}{0}
\setcounter{figure}{0}
\setcounter{section}{0}
\renewcommand{\thesection}{S\arabic{section}}
\renewcommand{\theequation}{S\arabic{equation}}
\renewcommand{\thefigure}{S\arabic{figure}}
\renewcommand{\cite}[1]{\citep{#1}}

\section{Further properties of exceptional bound (EB) spectra}
In this supplementary section, we provide further plots of various aspects of EB spectra.
\begin{figure*}[h]
    \centering
    \subfloat[]{\includegraphics[width = .3\linewidth]{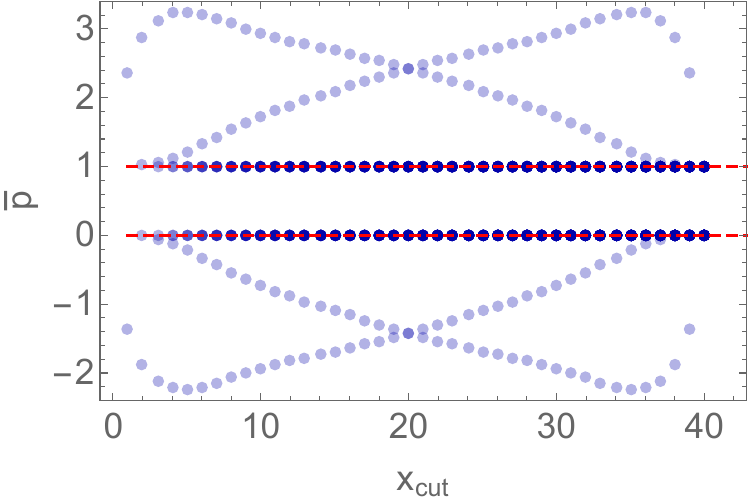}}
		\subfloat[]{\includegraphics[width = .3\linewidth]{xcutB3_a1.pdf}}
		\subfloat[]{\includegraphics[width = .3\linewidth]{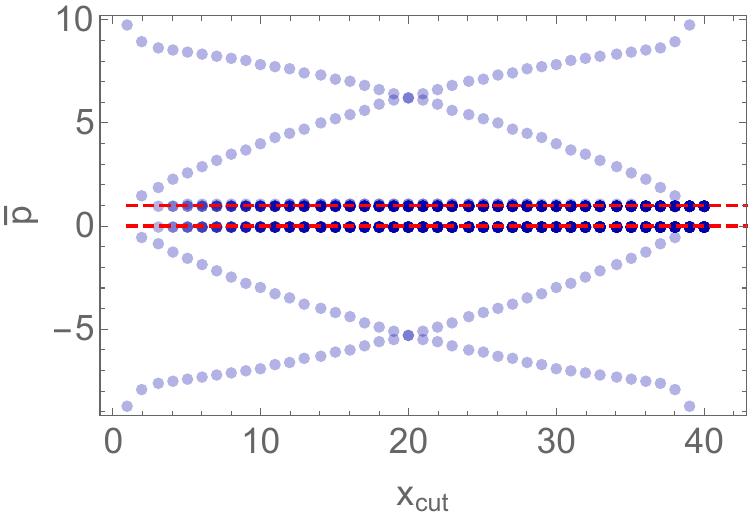}}
    \caption{\textbf{Effect of the magnitude of exceptional point asymmetry $a_0$.} Here we show how the EB spectrum $\bar p$ is affected by $a_0$ in the parent defective Hamiltonian $H_{n=2}(k)=h(k)\sigma_x + a_0(\sigma_x+i\sigma_y)$, $h(k)=\frac1{2}(2(1-\cos k))^B$, $B=3$. Depicted are (a) $a_0=0.01$, (b) $a_0=1$ and (c) $a_0=100$, all computed in a system with $L=40$ unit cells. As evident, the spectrum does not qualitatively change much, even though $a_0$ ranges over four orders of magnitude.
		}
    \label{fig:a0}
\end{figure*}

\begin{figure*}[h]
    \centering
		\begin{minipage}{.73\linewidth}
		\subfloat[]{\includegraphics[width = .49\linewidth]{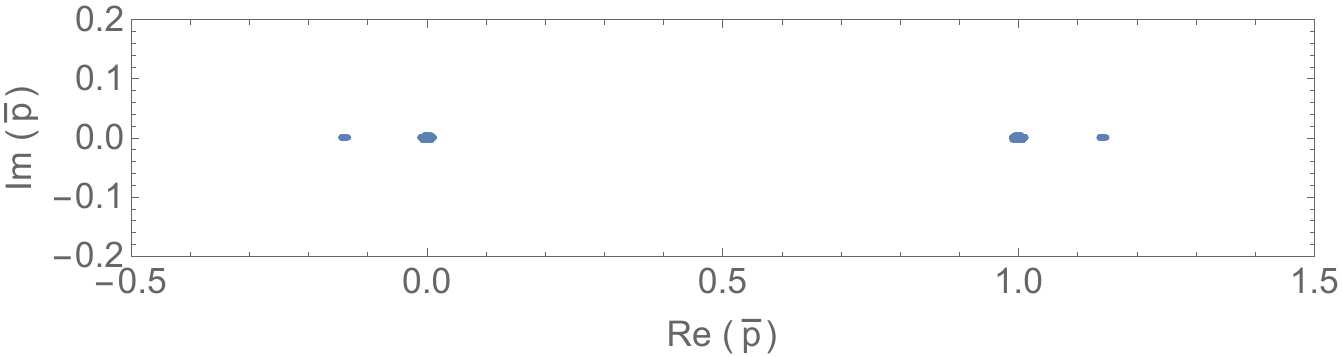}}
		\subfloat[]{\includegraphics[width = .49\linewidth]{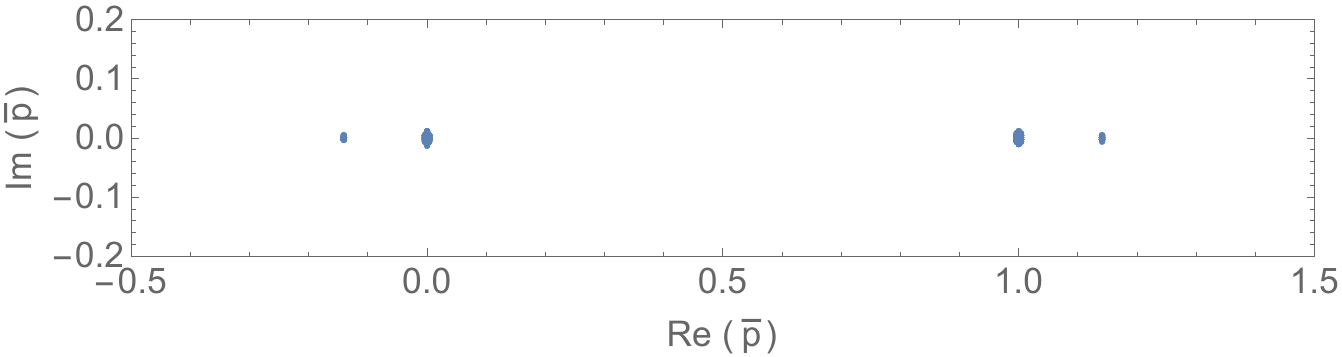}}\\
		\subfloat[]{\includegraphics[width = .49\linewidth]{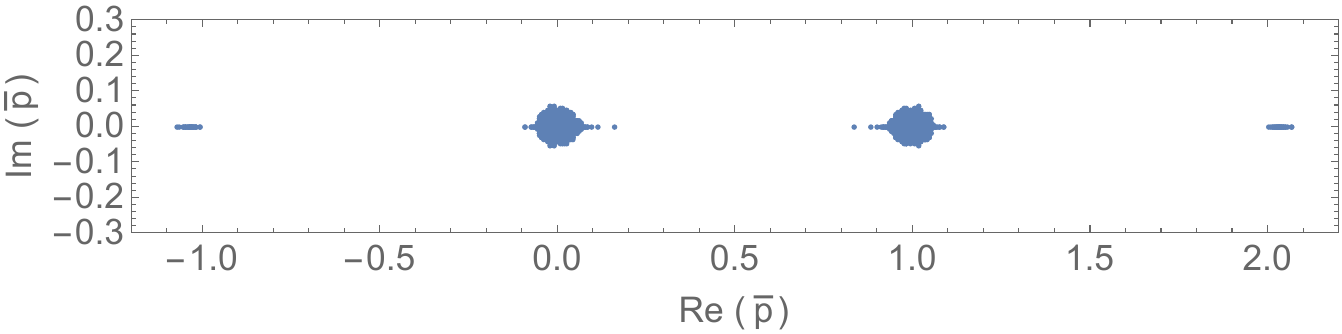}}
		\subfloat[]{\includegraphics[width = .49\linewidth]{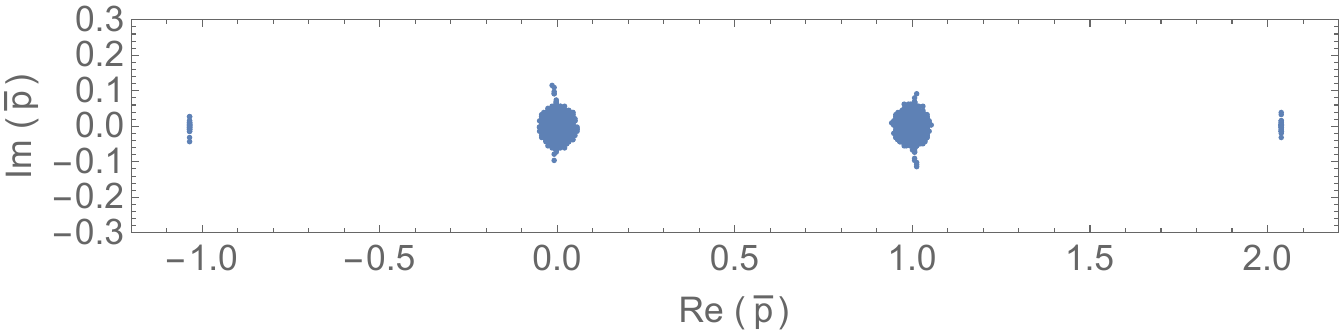}}\\
		\subfloat[]{\includegraphics[width = .49\linewidth]{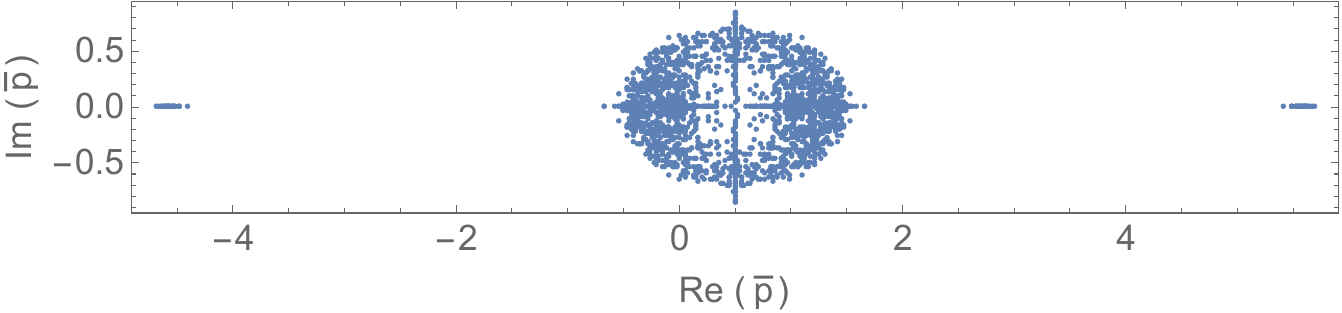}}
		\subfloat[]{\includegraphics[width = .49\linewidth]{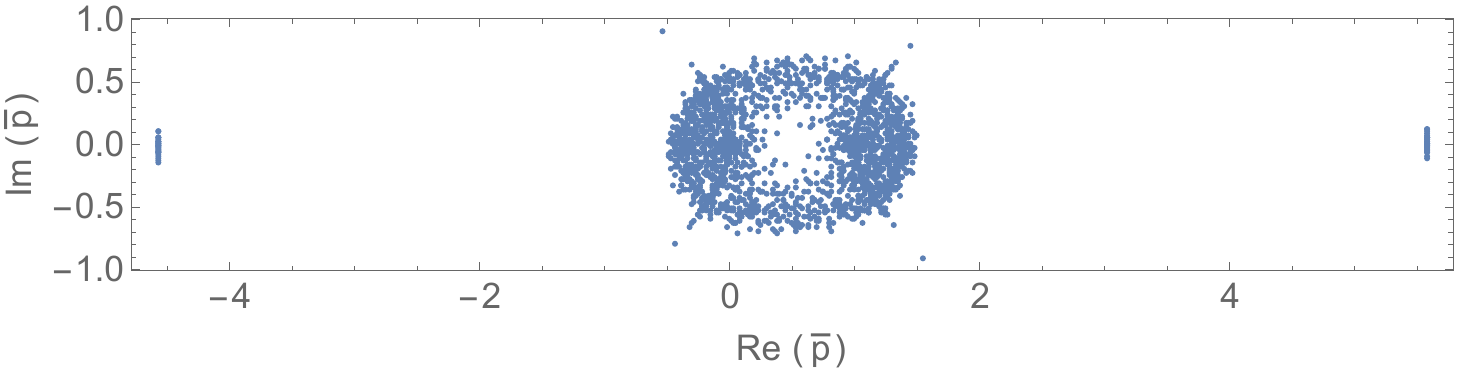}}\\
		\subfloat[]{\includegraphics[width = .49\linewidth]{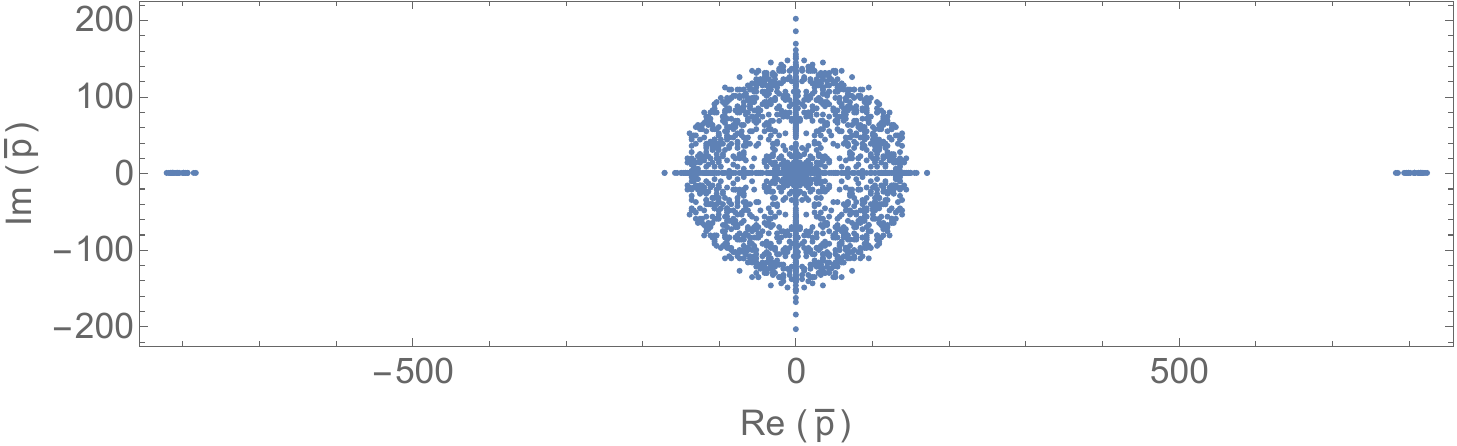}}
		\subfloat[]{\includegraphics[width = .49\linewidth]{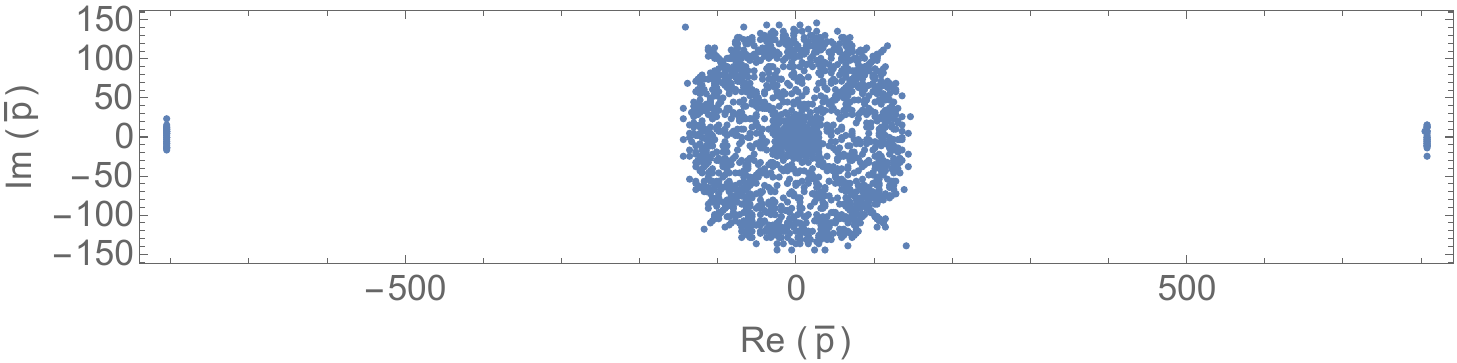}}
		\end{minipage}
    \caption{\textbf{Robustness of the EB spectrum with respect to real and imaginary disorder.} Distribution of eigenvalues $\bar p$ from 30 instances of random geometric disorder in the $\bar P$ matrix elements corresponding to $H_{n=2}$, with $L=40$, $x_{cut}=39$ and fractional disorder between $-1\%$ to $1\%$. The left column plots (a,c,e,g) depict real disorder while the right column plots (b,d,f,g) depict imaginary disorder. Each row depicts a different value of $B$: (a,b) $B=1$, (c,d) $B=2$, (e,f) $B=3$ and (g,h) $B=7$. The EB eigenvalues (far left or right) remain relatively unperturbed and well-resolved compared to the non-EB eigenvalues (central clouds), stays real/imaginary upon real/imaginary disorder, and in general also move further from $[0,1]$ with increasing $B$.
		}
    \label{fig:disorder}
\end{figure*}

\end{document}